\documentclass[final,5p,times,twocolumn,authoryear]{elsarticle}

\usepackage{framed,multirow}
\usepackage{subcaption}
\usepackage{amssymb}
\usepackage{latexsym}
\usepackage{geometry}

\usepackage{booktabs}
\usepackage{graphicx}
\usepackage{lscape}
\usepackage{adjustbox}
\usepackage{comment}
\usepackage{subcaption}
\usepackage[ruled]{algorithm2e}
\usepackage{graphicx}
\usepackage{adjustbox} 
\usepackage{amsmath}
\usepackage{gensymb}
\usepackage{textcomp}
\usepackage{siunitx}
\usepackage[table,xcdraw]{xcolor}
\SetAlFnt{\scriptsize}
\usepackage{xcolor}
\usepackage{multirow}
\usepackage{url}
\usepackage{xcolor}

\definecolor{newcolor}{rgb}{.8,.349,.1}
\usepackage{natbib}
\usepackage{hyperref}
\usepackage[switch,pagewise]{lineno} 

\begin{document}


\begin{frontmatter}

\title{\textbf{Predicting Geometric Errors and Failures in Additive Manufacturing\tnotetext[1]{Published in Rapid Prototyping Journal - cite as Ntousia, M., Fudos, I., Moschopoulos, S. and Stamati, V. (2023), "Predicting geometric errors and failures in additive manufacturing", Rapid Prototyping Journal, Vol. ahead-of-print No. ahead-of-print. https://doi.org/10.1108/RPJ-11-2022-0402}}}

\author[1]{Margarita Ntousia}
\author[1]{Ioannis Fudos} 
\author[1]{Spyridon Moschopoulos}
\author[1]{Vasiliki Stamati\corref{cor1}}

\address[1]{Dept of Computer Science and Engineering, University of Ioannina, 45110 Ioannina, Greece}


\cortext[cor1]{Corresponding author:}
\emailauthor{vstamati@cs.uoi.gr}{Vasiliki Stamati}

\begin{abstract}

\noindent\textbf{Purpose:} Objects fabricated using Additive Manufacturing (AM) technologies often suffer from dimensional accuracy issues and other part-specific problems. We present a framework for estimating the {\em printability} of a CAD model that expresses the probability that the model is fabricated correctly via an AM technology for a specific application.\\
\textbf{Approach:} We predict the dimensional deviations of the manufactured object per vertex and per part using a machine learning approach. The input to the error prediction artificial neural network (ANN) is per vertex information extracted from the mesh of the model to be manufactured. The output of the ANN is the estimated average per vertex error for the fabricated object.

 This error is then used along with other global and per part information in a framework for estimating the printability of the model, that is, the probability of being fabricated correctly on a certain AM technology, for a specific application domain. \\ 
\textbf{Findings:} A thorough experimental evaluation was conducted on binder jetting technology for both the error prediction approach and the printability estimation framework. \\
\textbf{Originality:} We present a method for predicting dimensional errors with high accuracy and a completely novel approach for estimating the probability of a CAD model to be fabricated without significant failures or errors that make it inappropriate for a specific application. \\
\textbf{Funding Statement:} This research was co-financed by the European Union and Greek national funds through the Operational Program Competitiveness, Entrepreneurship and Innovation, under the call RESEARCH -CREATE -INNOVATE (project code: T1EDK-04928).\\
\textbf{Ethical Compliance:} All procedures performed in studies involving human participants were in accordance with the ethical standards of the institutional and/or national research committee and with the 1964 Helsinki Declaration and its later amendments or comparable ethical standards.

\end{abstract}


\end{frontmatter}


\
\section{Introduction}
\label{sec:printability}
Manufacturing a product using Additive Manufacturing (AM) is a complex process that begins with the expression of design intent \citep{Kyratzi2020} and functionality by creating a CAD model and ends with the actual fabricated part \citep{Livesu2017}. Owing to several factors, such as the technical specifications of AM technologies, materials used, original CAD model design characteristics, meshing and slicing algorithms, the final fabricated parts may vary, sometimes significantly, from the original CAD models \citep{NGO2018}. In addition, fabricated parts often suffer from issues and part specific problems that affect the robustness, functionality and cost effectiveness of the manufacturing process. Therefore, there is a definite need for quality assurance processes in additive manufacturing \citep{Kim2018}.

\begin{figure*}[htb]
\centering
\includegraphics[width=0.9\textwidth]{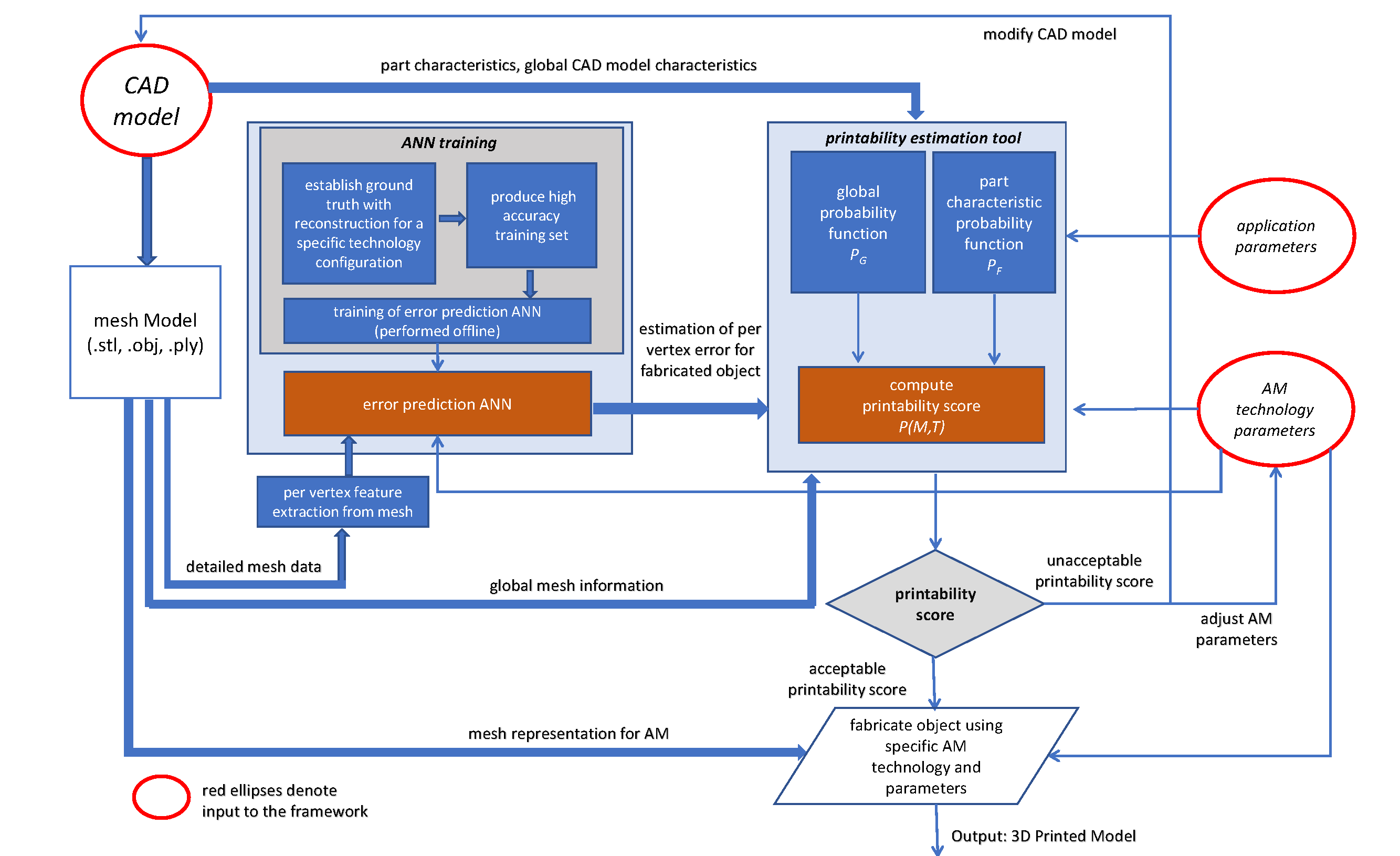}
\caption{Information flow diagram of the proposed framework (figure by authors).}
\label{fig:workflow}
\end{figure*}

In this study, we introduce the framework illustrated in Figure \ref{fig:workflow} for predicting printing failures and errors for quality assurance in AM processes. The input to the workflow consists of (i) the initial 3D CAD model which is to be manufactured, (ii) the additive manufacturing technology parameters and (iii) the application domain which are highlighted in red in Figure \ref{fig:workflow}.

The information extracted from the CAD model is used by the "error prediction ANN" of Figure \ref{fig:workflow} for predicting the dimensional deviations of the manufactured object (on a specific AM technology), per vertex, and per part.  The input to the ANN is per vertex information extracted from the mesh of the model to be manufactured and the output is the estimation of the average per vertex error for the fabricated object. The ANN is trained offline using a high-accuracy training set derived using a reconstruction method of the manufactured object to establish the ground truth for error estimation.

The error predicted by the ANN (output of the ANN) is then used as an input parameter in the process of estimating the probability (also known as \textit{printability}) of the model to be fabricated correctly on a certain AM technology, for a specific application (illustrated as \textit{"printability estimation tool”} in Figure \ref{fig:workflow}). Printability is computed by considering two factors: \textit{the global probability function}, which depends on the attributes and properties of the AM process used for manufacturing, for a specific application, and \textit{the part characteristic probability function}, which expresses the probability that a specific part  of the mesh model with certain characteristics will fail to print correctly.

As illustrated in the workflow of Figure \ref{fig:workflow}, if the estimated overall printability score is acceptable for the user, for a specific application, then the extracted mesh is recommended for fabrication under the specified parameter settings and the final output of the workflow is the 3D printed object. If not, then the user either: modifies the design characteristics of the initial CAD model, if the part characteristic probability function of a characteristic returns a high value; or alters the AM technology parameters, if the global probability function value affects the printability.

In a nutshell, the technical contributions of this paper are the following:
\begin{itemize}
    \item  A highly accurate reconstruction method for the final fabricated part using data acquired during the AM process. This method is more accurate than 3D scanners and much faster and more massive than other Coordinate-Measuring Machines (CMMs).
    \item A dimensional error prediction approach that is realized by an ANN that is trained using highly accurate data sets (source code for the pre-trained predictor and the feature extraction process is provided).
    \item A framework for estimating the probability of a model to be printed without failures (overall printability score), which depends on the AM technology to be used, application domain, predicted dimensional error and geometric design characteristics of the model to be manufactured.
    \item An easy to use interactive tool for determining the printability of a CAD model (source code is provided).
\end{itemize}

The remainder of this paper is organized as follows. Section 2 presents an overview of the literature on quality assurance, robustness and failure prediction. Section 3 presents an approach to predict the dimensional accuracy of a manufactured product, per vertex and per part, using machine learning methods. Section 4 focuses on the estimation of the probability that the model is printed correctly, on a specific AM technology. Furthermore, Section 5 presents validation experiments for the presented framework. Finally, Section 6 offers conclusions and future research directions.

 \section{Related work}
 Printing failures can result from either a single factor or a combination of many different ones. For example, it could be attributed to a low quality STL mesh, a CAD model design flaw, AM machine limitations, especially in reference to a specific model application domain, or a combination of the above. A geometric analysis of the printability of STL meshes is presented in \citep{Attene2016}, where an algorithm is also proposed for automatically repairing "unprintable" STL meshes. The algorithm analyzes the types of faults in the initial STL file and performs corrections based on triangle clusters. In the context of our study, we assume that the model mesh to be fabricated is robust and of high quality, requiring no corrections.    

Printing failures often occur owing to the design characteristics of the original model in relation to AM technology selected for fabrication. The quality of a manufactured part is directly linked to the design principles and rules followed during the design phase, prior to meshing and printing. There is an abundance of work that evaluates AM processes and correlates them with {\em design for additive manufacturing} (dfAM), see e.g. \citep{Booth2016}. 

Design rules have been determined for specific 3D printing technologies, to help designers produce high quality parts in conformance with the initial design. \citep{Mani2017} provided a survey of design principles that have been realized through a set of design rules for AM and a Guide-to-Principle-to-Rule (GPR) approach was proposed to assist the design process. Design rules are grouped together by \citep{Jee2015DESIGNRW} to form modules as a more dynamic and designer-friendly way of dealing with the design process and a case study is presented on powder bed fusion technology. Design rules for AM were developed and reported by \citep{Adam2015} based on geometrical standards and attributes that characterize the object shape.

\citep{Raffaeli2021} provide a procedural approach to help designers correlate the design process with the difficult decision of choosing a manufacturing method, if any is applicable. \citep{Zhang2016} presented an analysis of the design features for AM and a feature-based approach to determine the layering orientation for optimized additive manufacturing. \citep{Booth2016} addressed print failures. The authors designed and implemented a dfAM worksheet that was used in either in conceptual or CAD phase of the design process. The use of the worksheet has led to a decrease in print failures. \citep{Budinoff2021} presented a MATLAB-based virtual prototyping tool for helping and training users (especially novices) in good practices for designing models for AM manufacturing with Fused Filament Fabrication (FFF) technology. This tool analyzes part geometry and detects problematic geometries or orientation problems and makes suggestions to the user for better printing results.

In our work we studied and utilized all design rules, heuristics and guidelines, and integrated them into the printability estimation framework along with novel experimentally verified global and part characteristics.

The dimensional accuracy of various AM technologies is also a topic of considerable interest, because geometric deviations in the end product directly affect the overall quality. Dimensional accuracy refers to the extent to which a printed object matches the size and specifications of the original file. Studies have been carried out on technologies such as FDM \citep{HANON2021,Gulanova2020}, Powder Bed Fusion, and Binder Jetting (BJ) using different materials. Various methods have been proposed for predicting the dimensional accuracy of manufactured parts for quality control and assurance purposes. 

Quality assurance can be achieved by employing supervised learning methods to predict the quality of a manufactured item. \citep{Decker2021} analytically present research on shape deformation through predictive modeling and compensation approaches, dividing them into two main categories of predictive modeling approaches: physics-based approaches that utilize finite element modeling (e.g. \citep{Meier2021}) and data-driven approaches based on statistical analysis and machine learning \citep{ZHU2018}. In the scope of this study, we focus on statistical analysis and machine learning approaches owing to their more generalized outlook. Physics-based approaches are often restricted to specific AM technologies because of the specific physical phenomena they model in each case. 

\citep{DeckerHuang2019} introduced a shape-driven approach for predicting the geometric accuracy of an AM product based on machine learning that uses a set of eight predictor variables to express the geometric attributes of the triangular mesh model used for printing (.stl model). A novel method that trains a random forest to predict the geometric error of a vertex is presented. This model can then be employed to  predict the  dimensional deviations of other models, even for free-form design.  \citep{Decker2021} improved their previous method to achieve more accurate shape deviation prediction. This is an interesting first approach to error prediction that works well on small subsets of low-complexity objects. In our approach we used more input features (predictors) and trained a neural network with hundreds of thousands of examples, deriving better input-output correlation and accuracy, no over fitting and very good generalization. To this end, we proposed a pretrained neural network for per vertex geometric error prediction and source code for extracting features from an arbitrary mesh file.

\citep{Khanzadeh2017} presented an approach to quantify the geometric deviations of AM parts from a large data-set of laser-scanned coordinates using a self-organizing map (SOM). The aim of the work is to substantiate a relationship between the AM process conditions and the geometric accuracy of the end result. The results are demonstrated using FFF technology. 

\citep{Shen2019} addressed the problem of automatic error compensation. A framework is presented in which data obtained by methods such as 3D scanners and other Coordinate Measuring Machines (CMMs) are used to train a neural network that approximates the deformation function that describes how the input is deformed into an actual printed object. Subsequently, the inverse deformation function was approximated by using the output model as the training set to obtain the input model. \citep{He2018} presented a similar approach, where a deep learning method using a convolutional encoder-decoder network was presented to simulate the shape deformation of the P-$\mu$SLA process. These methods have the advantage of considering neighboring vertices when determining the deformation function of a vertex. However, they are slow and their generalization is not substantiated because they have been tested with simulated models, not with actual manufacturing data. Our error prediction method uses only actual data derived from high precision reconstruction of the manufactured object. The reconstruction method applied in this study is presented in the following section, where the ground truth is established. \citep{Wang2020} provide an overview of point cloud acquisition and processing methods  and compare different point cloud data acquisition approaches. Our method is highly accurate and uses a high resolution camera for data acquisition of per slice information. One of its strengths is that it is an automatic method that does not require user interaction, thereby eliminating human errors. Also, there is no significant error from noise, which is a common problem in laser scanning. Another advantage is that it can accurately capture very small parts and  the internal structures of the fabricated parts.

In our work we present a framework that combines the two different aspects of quality assurance and print failure/error prediction.
In reference to related work and our dimensional error prediction approach, our training data (output model) are obtained through a method of very high accuracy (more accurate than 3D scanners and much faster and more massive than other CMMs). In addition, we studied all plausible structural and geometric attributes of the vertices and determined features that were significantly correlated with the output error. Previous studies explored only a limited number of attributes. Our neural network error predictor works for meshes with arbitrary complexity. Furthermore, the proposed reconstruction method can be applied to any point cloud data captured using high accuracy methods. As for the printing failure probability estimation aspect of our work, the printability method can be applied for predicting the probability of failure for several AM technologies based on CAD model characteristics and assisted by the dimensional error prediction approach. Previous studies have considered specific characteristics and provided guidelines for designers. To the best of our knowledge there is no global framework for predicting the overall printability score \color{black}  of CAD models.  In the scope of this paper, our method is applied and evaluated for binder jetting technology.

\section{Error prediction}
\label{sec:prediction}
In this section we present an approach for accurately reconstructing the mesh model of a manufactured part during the fabrication process to determine the actual error of the fabricated object as compared to the original triangulated mesh provided as input to the 3D printer. This information is then used to train an ANN to predict the per vertex geometric error induced by the fabrication process. For reproducibility/repeatability purposes we provide both  a pretrained neural network and source code to derive the features from an arbitrary mesh file\footnote{Pretrained model and feature extraction: \\ 
\url{https://github.com/spirosmos/ANN_c2c_error_prediction}}.

\subsection{Establishing ground truth}

\label{sec:recon}

 Various methods can be used to derive the point cloud of manufactured objects. We introduce an automatic reconstruction process that derives point clouds with high accuracy during the fabrication process using a high resolution action camera that records the process. Other methods for 3D reconstruction (3D scanning, photogrammetry, etc.) can be used in AM technologies where this method cannot be applied (see e.g. \citep{eketaq3d}). In such cases, one should ensure high reconstruction accuracy for all parts of the object that will be used for training.

 To demonstrate  the reconstruction process we used the ZPrinter 450 3D printer that employs binder jetting technology. In this technology, gypsum powder is spread on the printing bed layer upon layer. Each time a layer of powder is spread, the printer prints the corresponding slice/layer of the model to be fabricated by jetting binder fluid and ink onto the powder. After a layer is printed, the printing bed is lowered, and another layer of powder is spread onto the bed. For all experiments, the layer thickness is set to 0.102 mm and a black binder cartridge was used without any additional coloring.

\begin{figure}[htbp]
 \centering
 \includegraphics[width=0.5\textwidth,trim={ 30mm 85mm 25mm 15mm },clip]{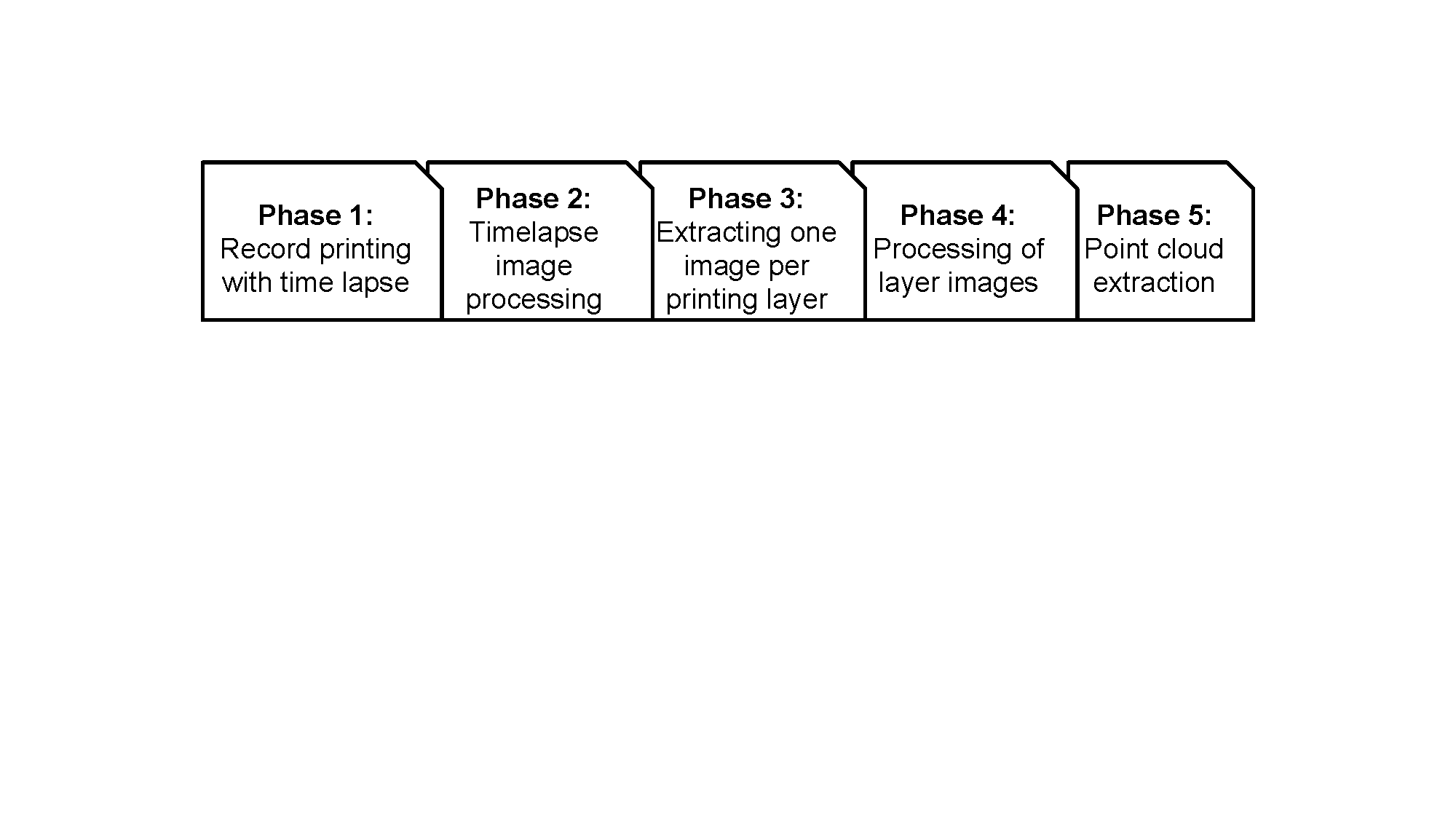}
 \caption{Workflow of data collection and processing(figure by authors).}
 \label{fig:steps}
\end{figure}

We developed a data collection and processing workflow (Figure \ref{fig:steps}) through which it is possible to reconstruct and visualize any model that is being printed in a fully automated manner with no user interaction.

\textit{Phase 1:} Initially, we record printing using a high resolution action camera placed in a vertical position above the printing bed. The camera records the time-lapse of the entire print, at maximum resolution (4K). To avoid additional distortions, the fish-eye effect is turned off. 

\textit{Phase 2:} When printing is complete, parts of the image that are not useful are cropped, and each frame is stored as an image file.

 This results in a sequence of thousands of images, for which we need to extract exactly one image for each printed layer.

\begin{figure}[htbp]
 \centering
 \includegraphics[width=0.8\columnwidth,]{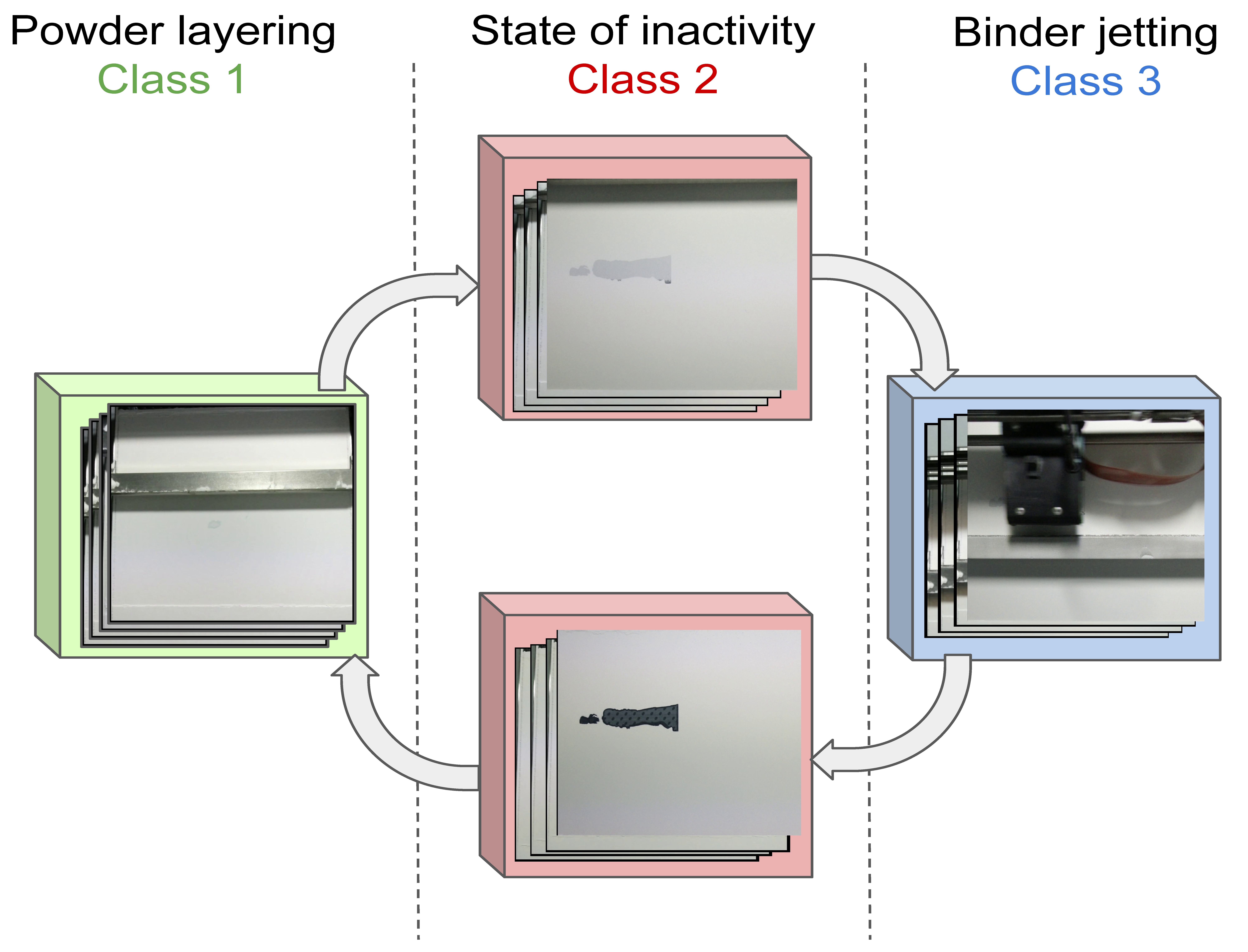}
 \caption{Printing steps, categorization scheme (figure by authors).}
 \label{fig:ClassSteps}
\end{figure}

\textit{Phase 3:}  A convolutional neural network (CNN) is used to classify the printing snapshots into distinct classes. We observed that ZPrinter 450 follows a process of repeated standard steps both before and after printing a new layer. This allowed us to classify the images that corresponded to each of the three steps to the respective class (Figure \ref{fig:ClassSteps}). Each step may contain  three to fifty images per layer.

The CNN is trained to classify images into three mutually exclusive classes with the highest possible accuracy. After classification of all images, we pick the first from each phase of inactivity (class 2) that follows a phase of binder jetting (class 3). This transition signifies the  completion of the binder jetting process for a new layer. The collection of printed layer images is then processed to reconstruct the point cloud of the printed solid. 

\begin{figure}[htbp]
 \centering
 \includegraphics[width=0.8\linewidth]{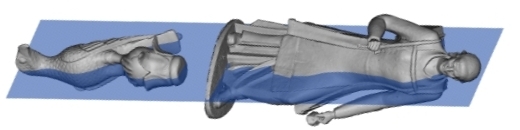}
 \includegraphics[width=0.8\linewidth]{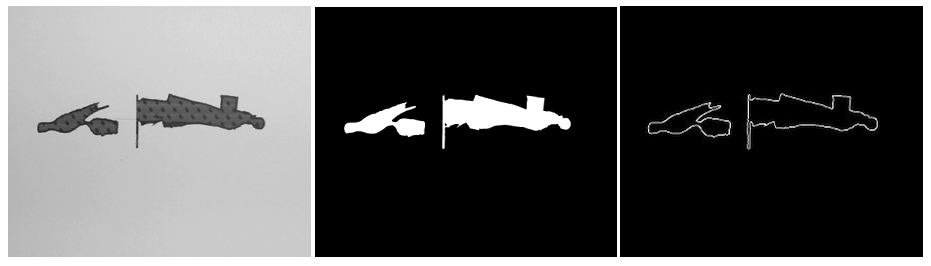}
 \caption{Steps of layer processing (figure by authors).}
 \label{fig:layer_processing}
\end{figure}

\textit{Phase 4:} We performed perspective correction and produced images with dimensions proportional to print level. Then, layer images were processed to have only black and white pixels. We converted images from RGB to grayscale and, with the use of a threshold in (0, 255), the pixels of every image are mapped to black or white, depending on their grayscale values. Edge extraction was carried out using a Canny edge extraction filter \citep{cannyFilter}. The result is an image with a black background and white lines with a thickness of 1 px which represent the detected edges, as shown in Figure \ref{fig:layer_processing}.

\textit{Phase 5:} We used the processed images derived in {\em Phase 4} to reconstruct the point cloud that corresponds to the fabricated part. According to the technical characteristics of the printer, the dimensions of the print bed are $254 mm \times 203 mm$. From these characteristics, the coordinate ranges are: $x\in[0, 254]$, $y\in[0, 203]$ and $z\in[0, L*S] mm$, where $L$ is the number of layers and $S$  the thickness of each layer. For the training described in this section, $S$ was set to $0.102 mm$. For each white pixel in a layer, we computed the corresponding coordinates $(x, y, z)$ of the point cloud slice.
 
\begin{figure}[htbp]
\centering
\resizebox{8.5cm}{!}{
\begin{tabular}{cccc}
      \includegraphics[height=30mm]{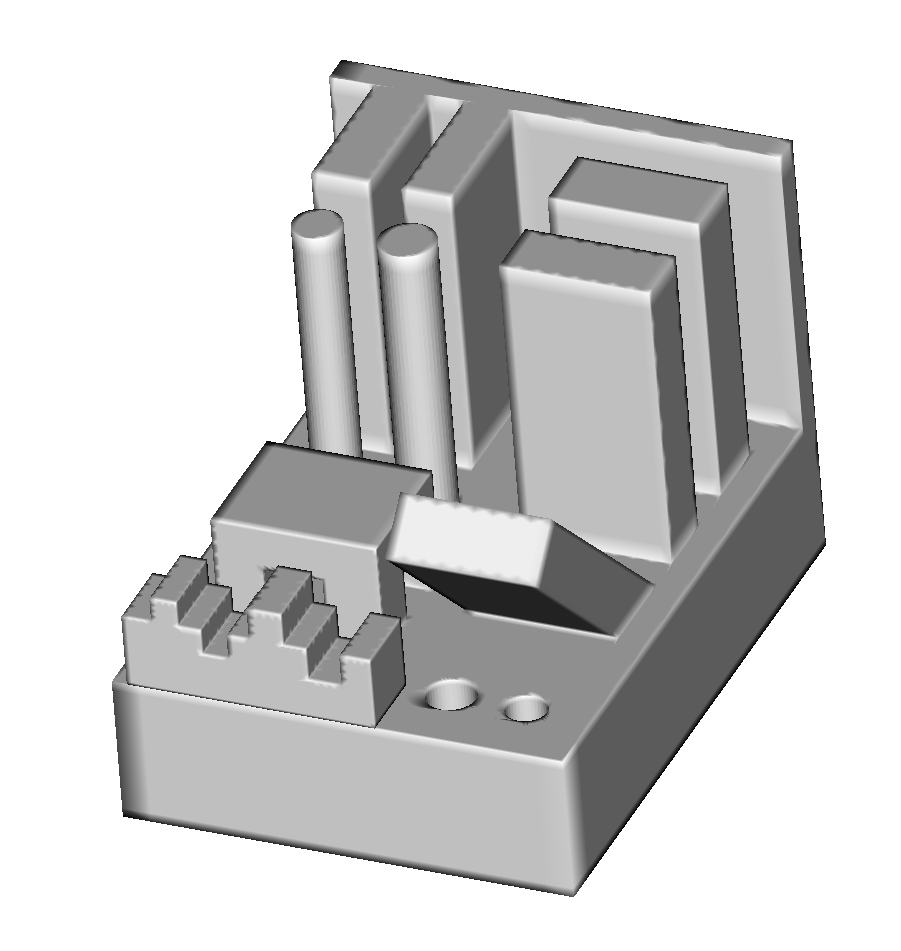} &
      \includegraphics[width=15mm]{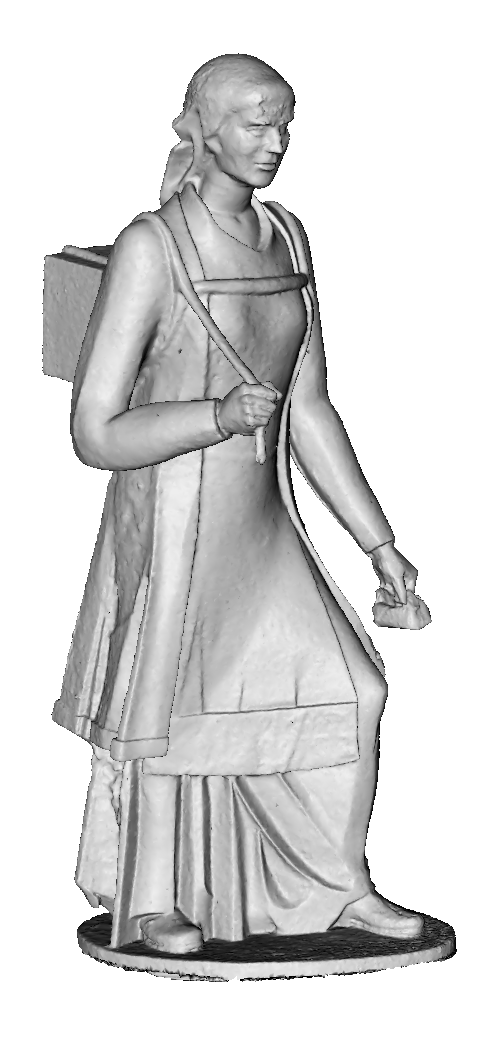} &
      \includegraphics[width=15mm]{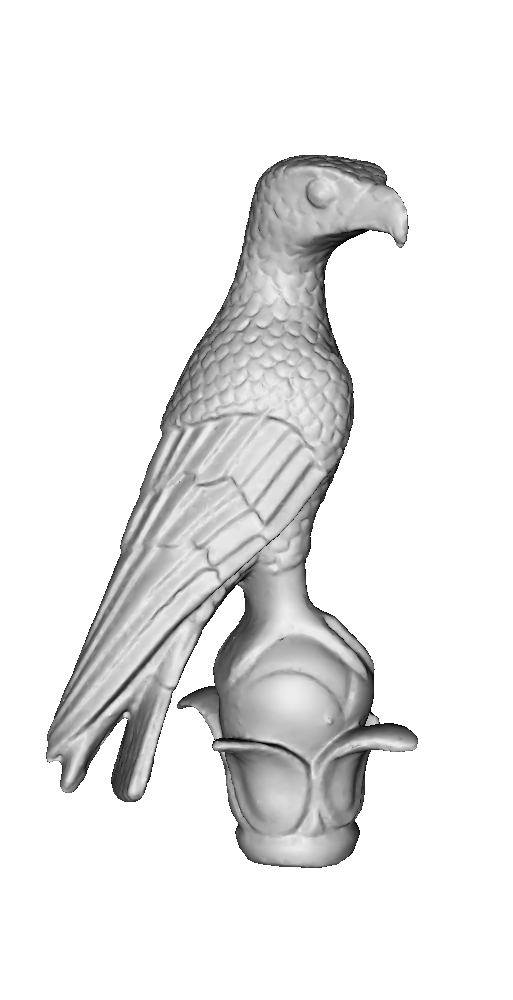} &
      \includegraphics[width=15mm]{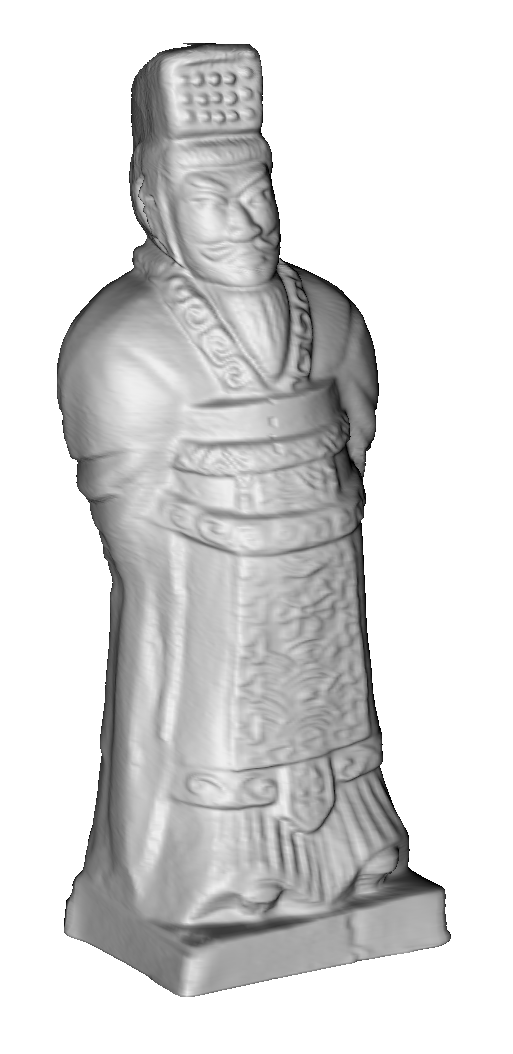}\\
      \includegraphics[height=30mm]{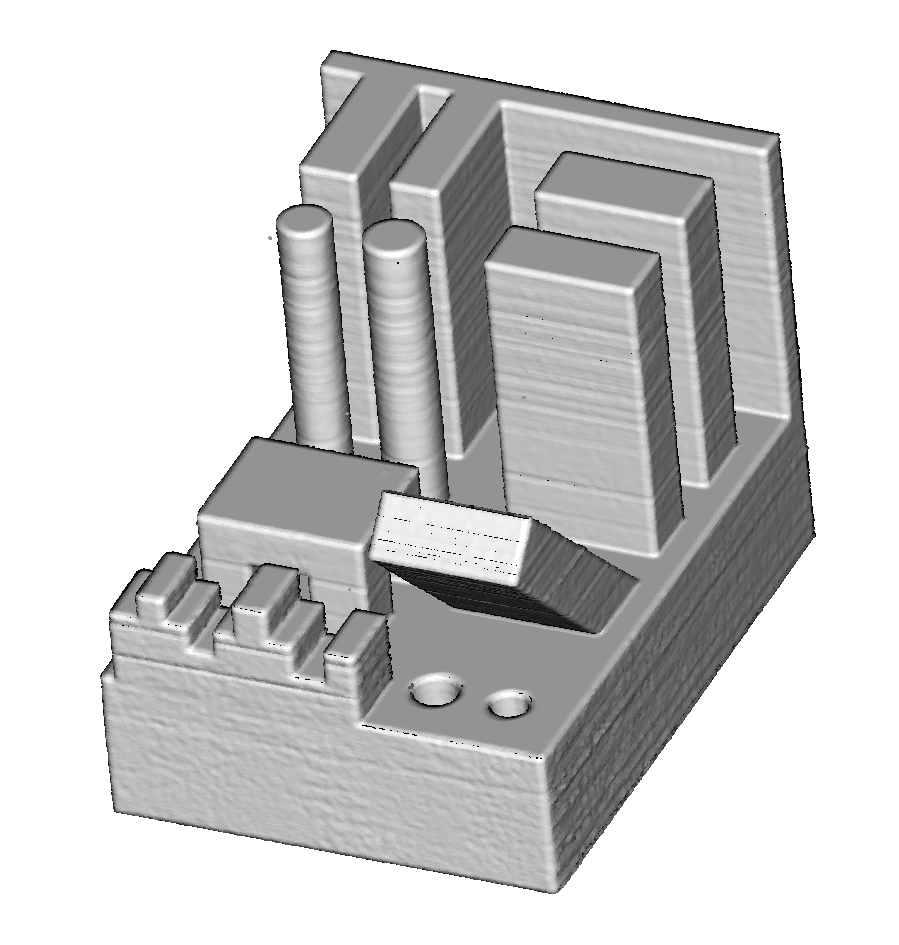} &
      \includegraphics[width=15mm]{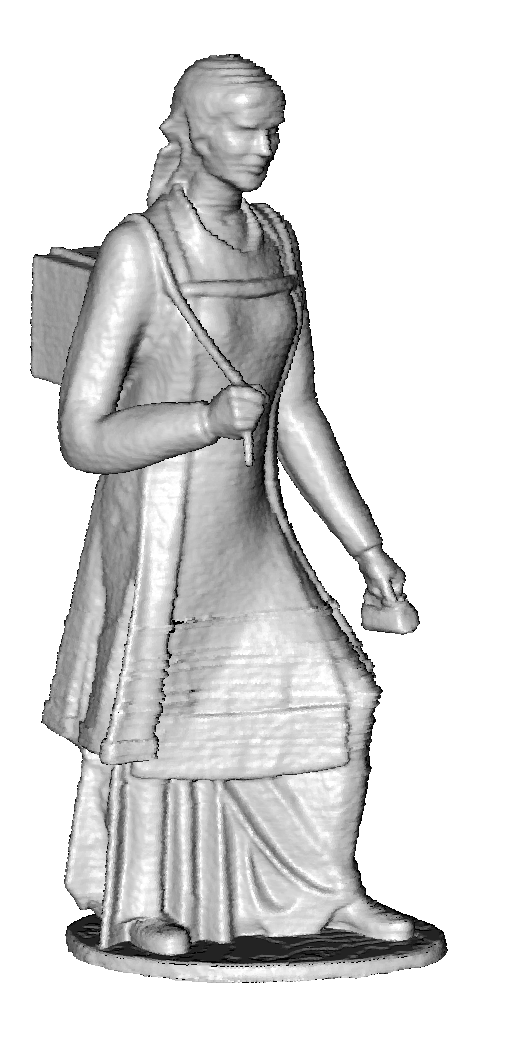} &
      \includegraphics[width=15mm]{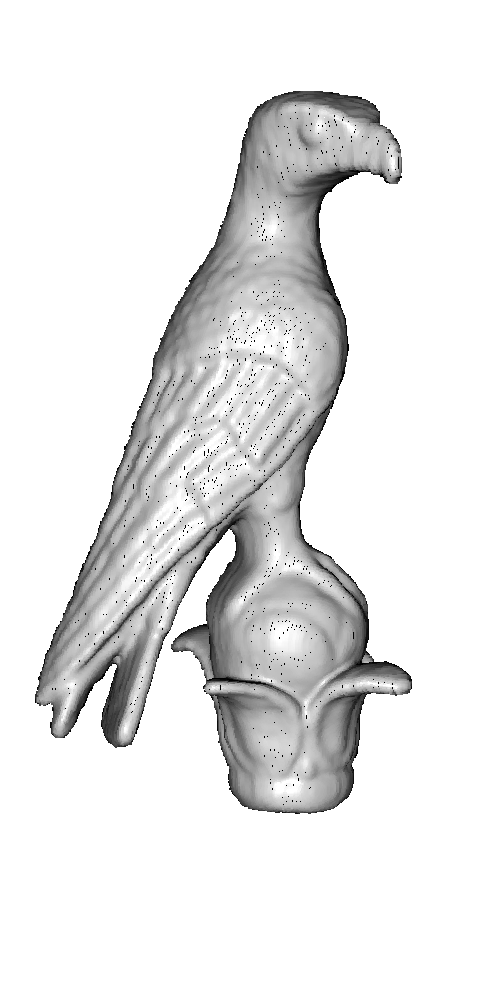} &
      \includegraphics[width=15mm]{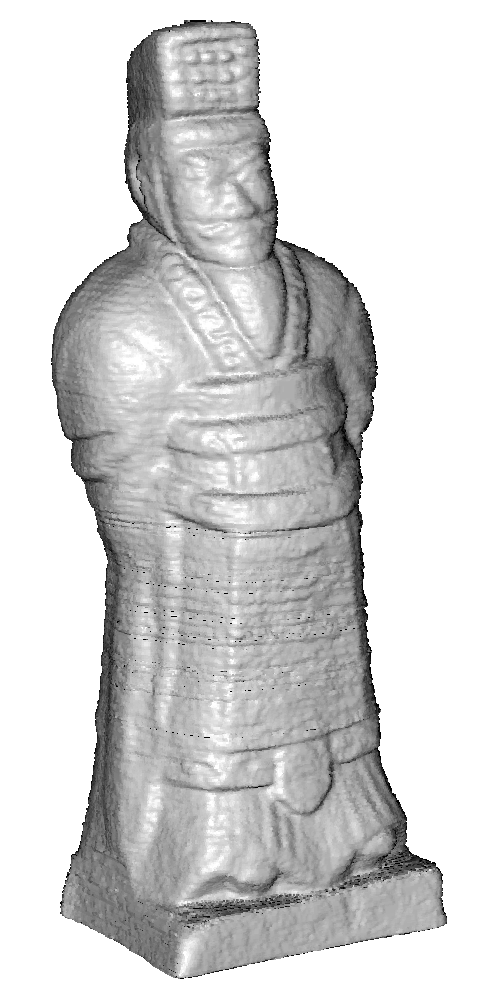}\\
\end{tabular}
}
\caption{Reconstruction examples: Original meshes (top), reconstructed point clouds (bottom) (figure by authors).}
\label{fig:rec_results}
\end{figure}

An issue is how to reconstruct faces parallel to the printing bed ($xy$ plane). To this end, we have also considered cross-sections that are parallel to the $xz$ plane. We then repeated the reconstruction process while constantly moving towards the $y$-axis. A high density point cloud, approximately $150 - 200 points/mm^{2}$, was exported in .ply file format. The file contains the coordinates of all points as well as the number of points. Examples of the initial meshes and reconstructed point clouds of the fabricated objects are shown in Figure \ref{fig:rec_results}. 

\begin{figure}
\centering
\includegraphics[width=0.75\linewidth]{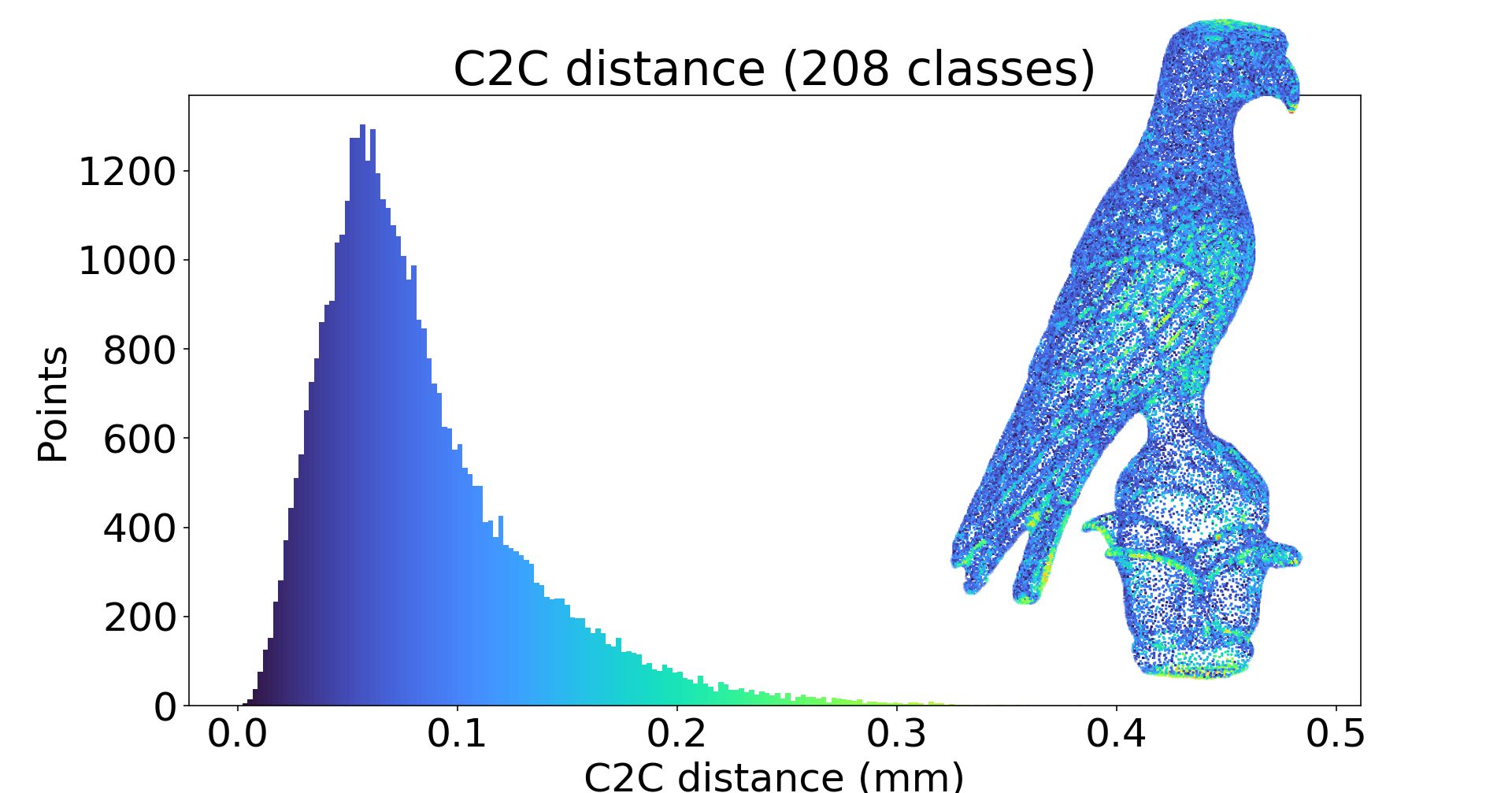}
\caption{C2C distance for the initial set of points (figure by authors).}
\label{fig:c2c_met}
\end{figure}

The next step of Phase 5 is to compute the dimensional accuracy of the original mesh compared to the reconstructed point cloud to determine the dimensional error introduced by the printing process. It should be noted that the reconstructed point cloud is assumed by definition to have the same scale as the input mesh. To compute the dimensional error we applied a 2-step procedure, during which, the scale factor was not adjusted. First, the point cloud of the original mesh was aligned with the reconstructed point cloud of the fabricated object. To this end, we used the iterative closest point (ICP) registration method \citep{C2C_ICP}. 
Second, the per vertex geometric error is computed using the cloud to cloud distance (C2C) \citep{C2C_ICP} (Figure \ref{fig:c2c_met}). Thus, we derived the absolute error (in millimeters) for each vertex of the input mesh according to its Euclidean distance from the reconstructed point cloud. Because the reconstructed point cloud of the fabricated object is very dense, we considered the minimal $L_2$ distance from all points on the reconstructed point cloud towards the point cloud of the original mesh.

\subsection{A neural network predictor for geometric errors in additive manufacturing}

We report on the development of an approach for predicting the per vertex geometric error of manufactured objects as compared to the mesh that is given as input to the 3D printer. A neural network is trained with samples from several object models using as features the vertex characteristics that exhibit a correlation with the geometric error in terms of the Euclidean distance from the manufactured object (i.e. how far from the correct position a vertex lies in the printed object).

In the following, we assume that the triangulated mesh that is given as input to the 3D printer is transformed into the coordinate system of the printer. The following features are used as inputs to the neural network to predict the resulting error corresponding to a vertex:
\begin{itemize}
\item[(i)] The positioning of the mesh in the printing area may have a significant impact on the accuracy. This has been reported for  positioning on a powder bed by several researchers (see e.g. \citep{Mostafaei2021,Farzadi2014}). Thus, we used the coordinates $(x, y, z)$ of each point as three separate input features for the ANN (features $F_1$, $F_2$ and $F_3$). These coordinates corresponded to the actual position of the vertex on the printing bed.

\item[(ii)] The Gaussian and mean curvatures of each vertex  provide information about surface complexity. For example, high curvature values often correlate with significant printing errors because the slicing direction may not be able to model abrupt changes in the surface. Therefore, these values were used as the fourth and fifth features of the ANN (features $F_4$ and $F_5$).  We used discrete operators introduced by \citep{Curvatures2003} because their estimates are optimal in terms of accuracy under mild smoothness conditions which can be enforced on triangulated meshes used for 3D printing. 

\item[(iii)] We define three more features that express the range of angles of triangles adjacent to any given vertex. Specifically, the maximum, minimum and average of all angles for that vertex, are used to capture information regarding the robustness of the triangular representation (features $F_6$, $F_7$ and $F_8$). Extreme angles are often correlated to the increased numerical errors incurred during the slicing process. The slicing process in many technologies is performed using proprietary slicer software.  

\item[(iv)] A significant feature is the angle between the normal vector of each point and the slicing direction which coincides with the z-axis (feature $F_9$).  This is probably the most well studied feature in the literature and several studies suggest that there is a high correlation between the error and the building direction (e.g. \citep{byun2006,pandley2017}).  

\item[(v)] Points near the bounding box often have a lower dimensional accuracy than other points. This is an observation that we have made and is probably caused by the controllers of the electro-mechanical parts of the 3D printers and the corresponding algorithms.  Hence, the minimum Euclidean distance of each vertex to the bounding box was used as the tenth feature of the input (feature $F_{10}$). 
\end{itemize}

Each of these characteristics plays an essential role in the predictive performance of a model. The removal of any feature increases the training and validation errors of the network. Overall, this set of features can be calculated for any triangulated model and provide a reliable estimate of the error of the fabricated item.

\subsection{Network architecture and evaluation}
\begin{figure}[htbp]
\centering
  \begin{subfigure}[b]{0.5\columnwidth}
  \centering
    \resizebox{3cm}{!}{
      \begin{tabular}{|c|}                              \cline{1-1}
        Input layer (10)                             \\ \cline{1-1}
        Dense (128), ReLU                            \\ \cline{1-1}
        Dense (128), ReLU                            \\ \cline{1-1}
        Dense (64), ReLU                             \\ \cline{1-1}
        Dense (1), ReLU                              \\ \cline{1-1}
        \end{tabular}
    }
    \caption{Artificial neural network architecture.}
    \label{fig:ann_ar}
  \end{subfigure}
  \begin{subfigure}[b]{\columnwidth}
  \centering
    \includegraphics[width= 0.8\linewidth]{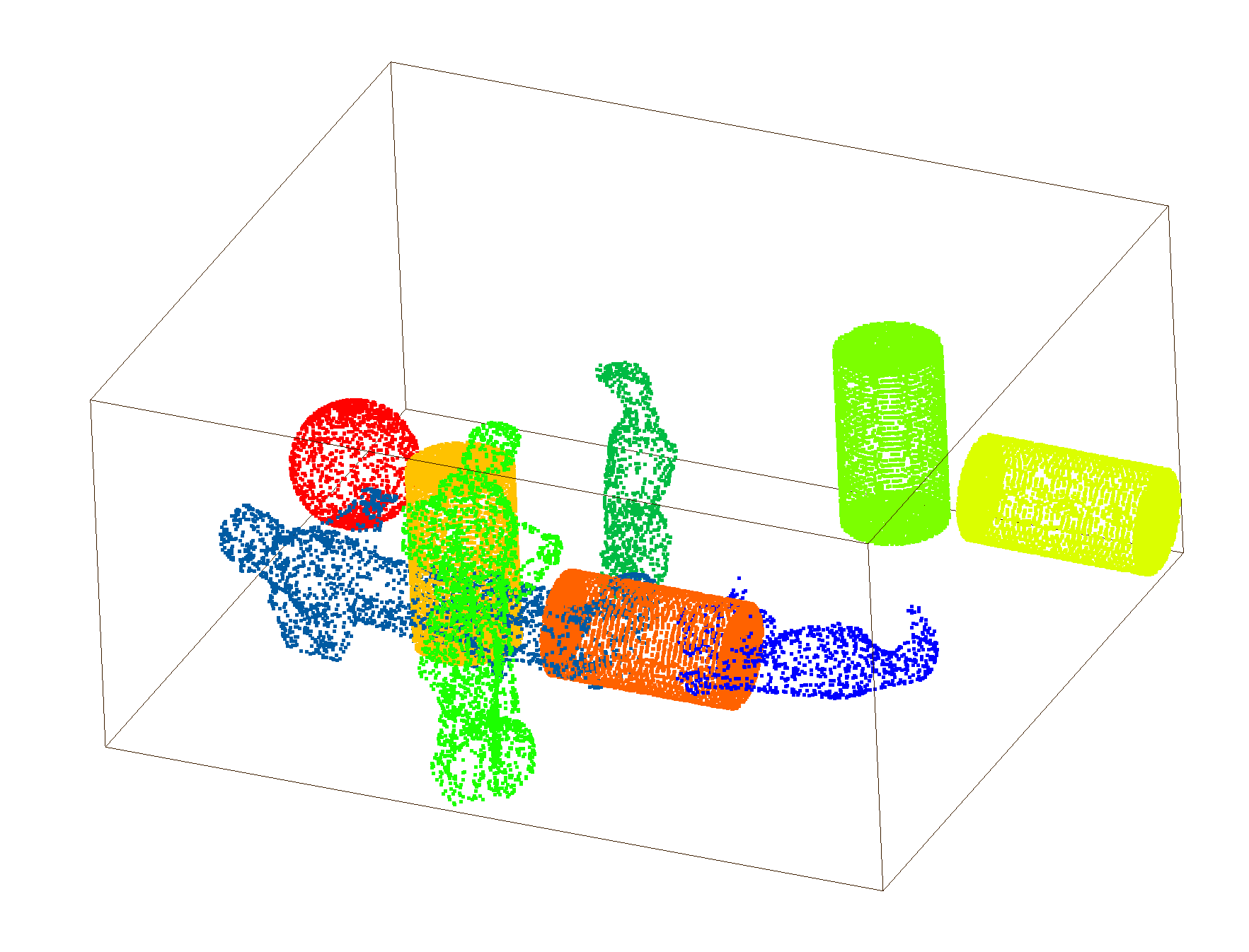}
    \caption{Training set.}
    \label{fig:trainingSet}
  \end{subfigure}
  \begin{subfigure}[b]{\columnwidth}
  \centering
    \includegraphics[width=0.7\linewidth]{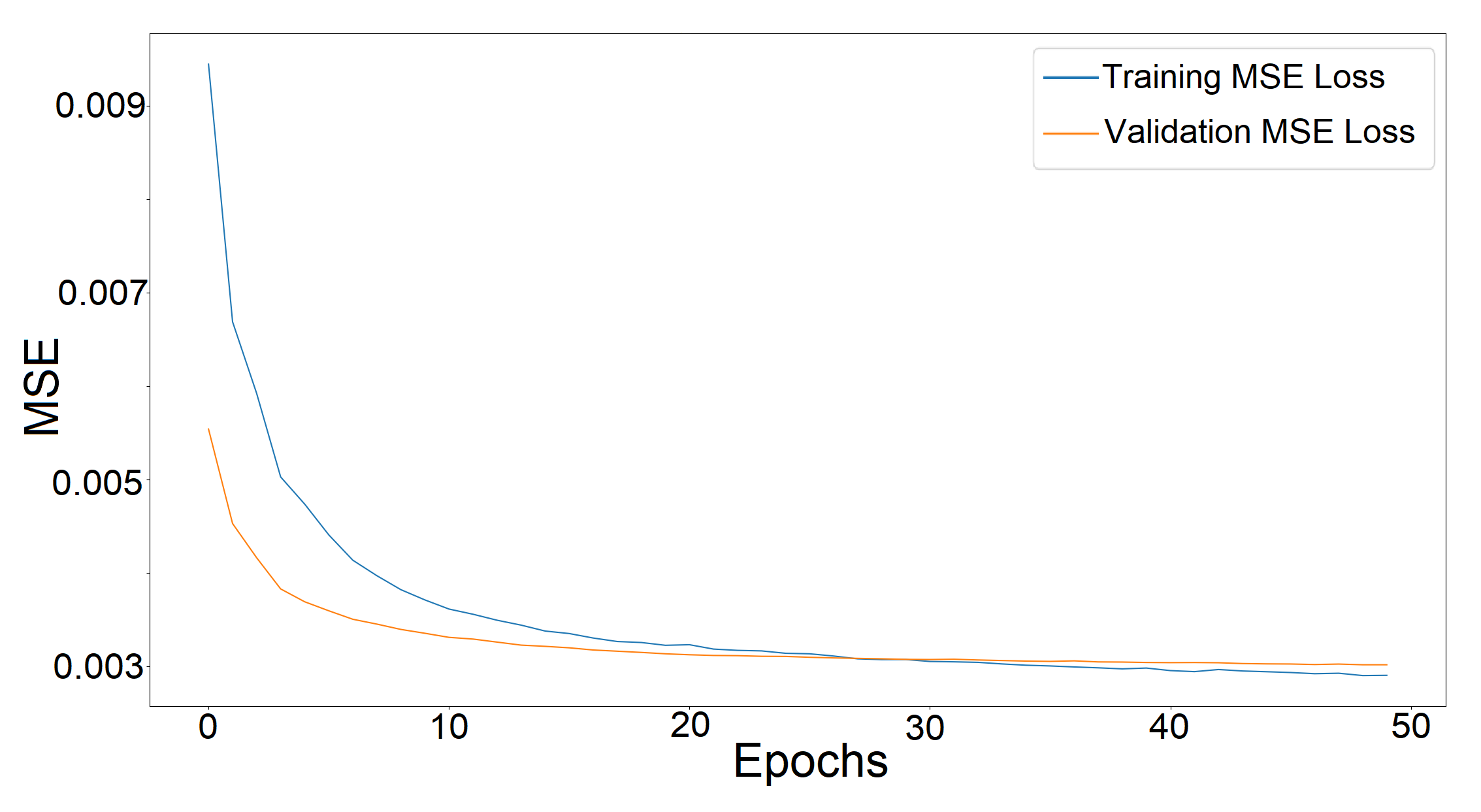}
    \caption{Training/Validation loss.}
    \label{fig:ann_loss}
  \end{subfigure}
\caption{Artificial neural network for error prediction (figure by authors).}
\label{fig:ann}
\end{figure}

The ANN utilizes a simple architecture consisting of an input layer, three hidden layers and an output layer (Figure \ref{fig:ann_ar}). The number of neurons in the hidden layers was determined by experimenting with several alternative networks. In addition, rectified linear units (ReLU) are used for every neuron of the network. We measure the loss of the ANN with the mean squared error loss function (MSE) and train the network with backpropagation for up to 50 epochs (Figure  \ref{fig:ann_loss}) using as training set subsets of the ``Dodone Eagle", ``Woman of Pindos", printed in horizontal and vertical orientation and five geometric primitives (four cylinders, one sphere) (Figure \ref{fig:trainingSet}).

\begin{figure}[htpb]
  \begin{subfigure}[c]{\columnwidth}
   \centering
    \includegraphics[width=0.7\linewidth]{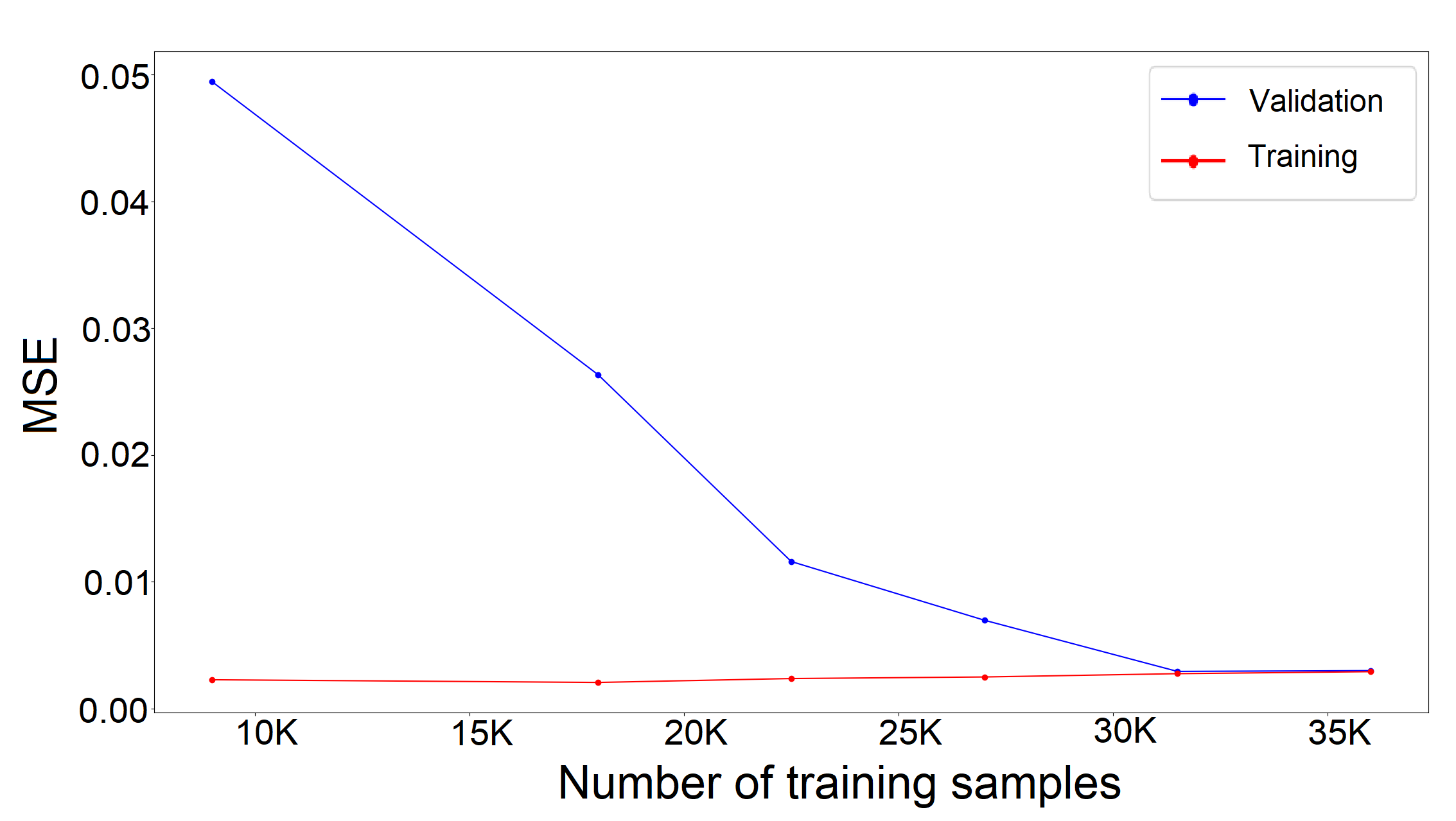}
    \caption{Training and validation loss versus size of the training set.}
    \label{fig:learningCurvesA}
  \end{subfigure}
  \begin{subfigure}[c]{\columnwidth}
  \centering
    \includegraphics[width=0.72\linewidth]{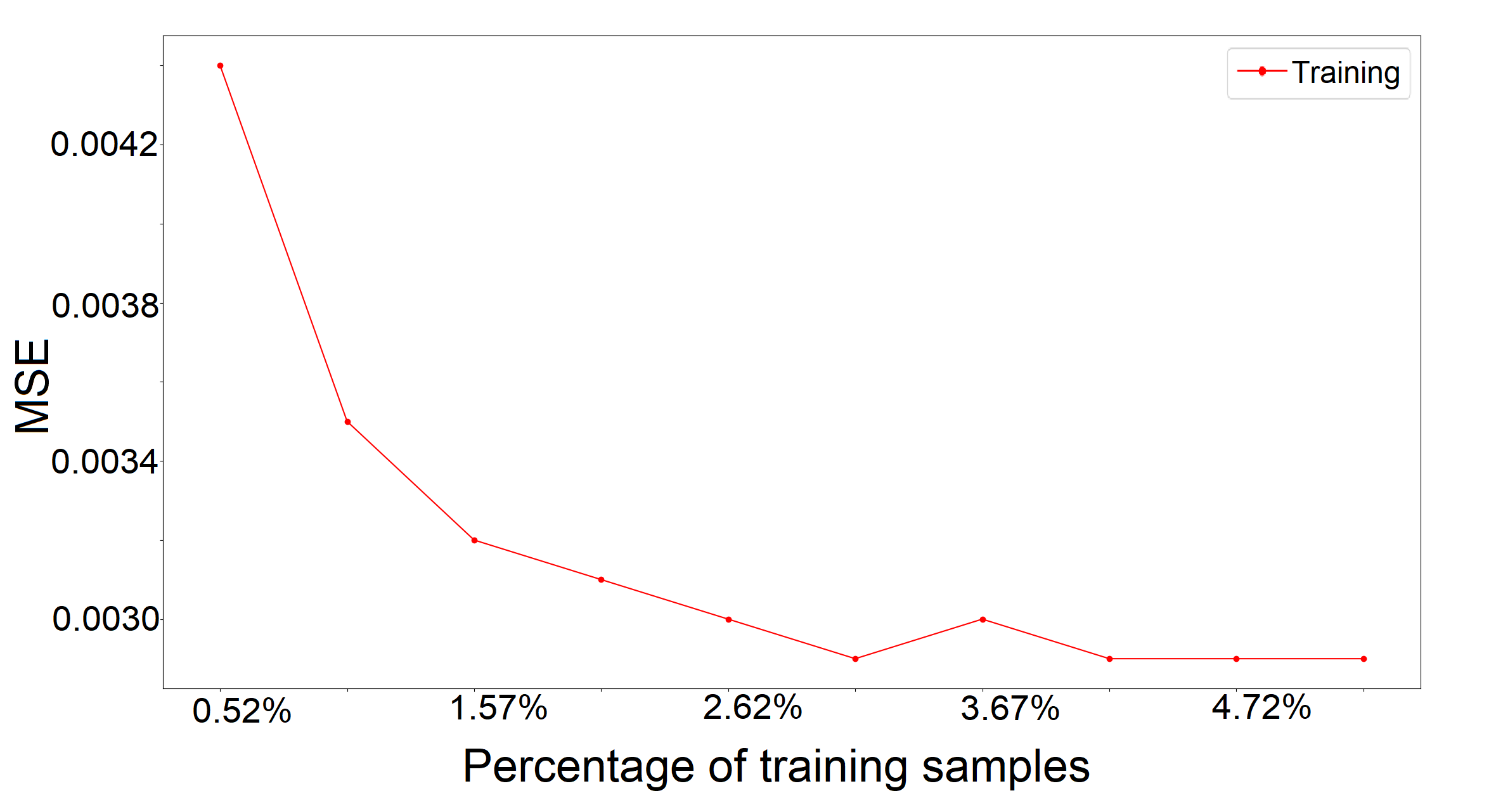}
    \caption{Training loss versus percentage of training data samples.}
    \label{fig:learningCurvesB}
  \end{subfigure}
\caption{{Learning curves (figure by authors).}}
\label{fig:learningCurves}
\end{figure}

To evaluate the network performance, we measured the training and validation loss values against the cardinality  of the training set, which was increased from 10k to 37.5k  by adding samples that are subsets of different mesh parts. Each subset differs in position on the printing bed, surface complexity, scale, and orientation along the z-axis. The results are illustrated in Figure \ref{fig:learningCurvesA} and indicate that both training and validation sets converged at a small mean square error. This is interpreted as a good fit capable of generalization.

The original point clouds have a density of $0.077\,mm$ to $0.278\,mm$. To derive the optimal size of the training set we trained the ANN using a uniformly sampled percentage of our data. We increased this percentage starting from $0.52\%$ and selected the minimum sample size for which the MSE converges (Figure \ref{fig:learningCurvesB}). This minimum percentage of sampling is determined to be in the area of $4.2\% - 4.5\%$. For example, if we have a model with a total number of 760k vertices, we deduce that a number of uniformly sampled vertices between 30.4k and 34.2k is expected to produce a well fitted model.

Our method produces representative results for the dimensional error of a vertex of a mesh and accurate results for the dimensional error for a characteristic of a part that is derived from the average predicted error of all the vertices of this part.

\section{Printability estimation framework}
\label{sec:printability general}
An important issue in AM is the ability to predict geometric errors and manufacturing failures that may occur when manufacturing a part. Such defects and errors are often due to the model design characteristics (complex surfaces, very thin parts, tolerances etc.), AM process parameters (layer height, building direction, part orientation, printing speed etc.) and AM technology limitations (i.e. robustness, dimensional accuracy etc.) \citep{Prabhakar2020, Lennert2021}. A preliminary approach to predicting the probability of successfully manufacturing a 3D model via AM was presented by \citep{Fudos2020}. In the present work, a complete substantiated framework for estimating the probability of successfully fabricating a CAD model using a certain AM technology for a specific application is presented. 

The purpose of this work is to help the designer avoid bad design choices by predicting the probability of failures occuring to the point where the manufactured part is inappropriate for the intended application.  

The proposed printability framework can be applied to various AM technologies. We provide a detailed explanation of all choices for printability estimation for Binder Jetting technology, while the required information and conclusions for other technologies, such as Fused Deposition Modeling (FDM) and Material Jetting (MJ), are also  provided. 

To compute {\em printability} we use two functions that express the probability of printing failure: the \emph{global probability function} - $P_{G}(C_{M},T,A)$ and the \emph{part characteristic probability function} - $P_{F}(i,d,T,A)$. 
$P_{G}(C_{M},T,A)$ expresses the probability that the overall quality of the fabricated object deteriorates to the point where the result is not acceptable. This is related to a) the geometric complexity of the model $C_{M}$ to be manufactured, b) the characteristics (i.e. accuracy, surface texture etc.) of the AM technology $T$ to be employed (FDM, MJ, BJ, etc) and c) the intended application $A$ (e.g. mechanical, artistic, bio-engineering etc.). 

The second function focuses on the local part characteristics (pins, holes, thin parts etc.), denoted by $i$ , of the model to be manufactured. Such part characteristics include a set of dimensional parameters (e.g. height, thickness, and width). We have also introduced a novel parameter $\epsilon$ that expresses the mean absolute value of the dimensional error on characteristic $i$ as predicted by the ANN in Section \ref{sec:prediction}. Thus, summarizing the overall probability formula of a model $M$, with $n$ part characteristics, to be successfully printed using a certain AM technology $T$ for a specific application $A$ is given by Equation \ref{eq:overall}.

\begin{equation}
\label{eq:overall}    
   P(M,T, A)= (1-P_{G}(C_{M},T,A)) \prod_{i=1}^{n} (1-P_{F}(i,d,T,A))
\end{equation}

where $1-P_{G}(C_{M},T,A)$ and $1- P_{F}(i,d,T,A)$ are the probabilities of obtaining a  fabricated object without significant flaws. 

The overall printability score of model $M$ on technology $T$ for application $A$ is given by Equation \ref{eq:overall_score}:
\begin{equation}
\label{eq:overall_score}  
   OPS(M,T,A)=100*P(M,T,A)
\end{equation}

\subsection{Deriving the global and the part characteristic probability functions}
\label{sec:printability parameters}
The global probability function $P_{G}(C_{M},T,A)$ expresses the impact that each AM technology has on the printing outcome, and is more of a quality parameter. \citep{Fudos2020} have studied the most significant global characteristics denoted by $x$ (accuracy, surface texture, support construction and various other abnormalities) for FDM, BJ and MJ AM technologies. These global characteristics have been rated based on their quality, as high, moderate and low for each AM technology by \citep{ntousia2019}. An initial defect score, $DS_{T_{Perfect}}(x)$, was then assigned to each one based on the technical specifications of each technology $T$ and the experimental technology assessment presented in \citep{Ramya2016} which expresses the probability of a characteristic $x$ to cause a significant problem in the printed part, for the highest mesh resolution. The defect scores were assigned with the following defect probability values: ``***"=0.01, ``**"=0.03, ``*"=0.05, indicating respectively low, moderate and high probability for defects. The values chosen for the scoring schema were determined experimentally.

In summary, the total global probability function $P_{G}$ of a model $M$, for a specific application $A$ on a certain AM technology $T$, is given by Equation~\ref{eq:global}.
\begin{equation}
\label{eq:global}
   P_{G}(C_{M},T,A) = 1-\prod_{x \in S} (1-(1-(1-DS_{T_{Perfect}}(x))\,QS_{C_{M}})\,k(x,A))
\end{equation} 

where $S$ is the set of global AM technology characteristics, $k(x,A) \in [0,1]$ is a factor that determines the sensitivity of application $A$ to characteristic $x$ and $QS_{C_{M}}$ is the ratio of the mesh surface area $Area(M)$ to the surface area $Area(O)$ of the original CAD model.

\begin{figure*}[htbp]
\centering
\includegraphics[scale=0.45]{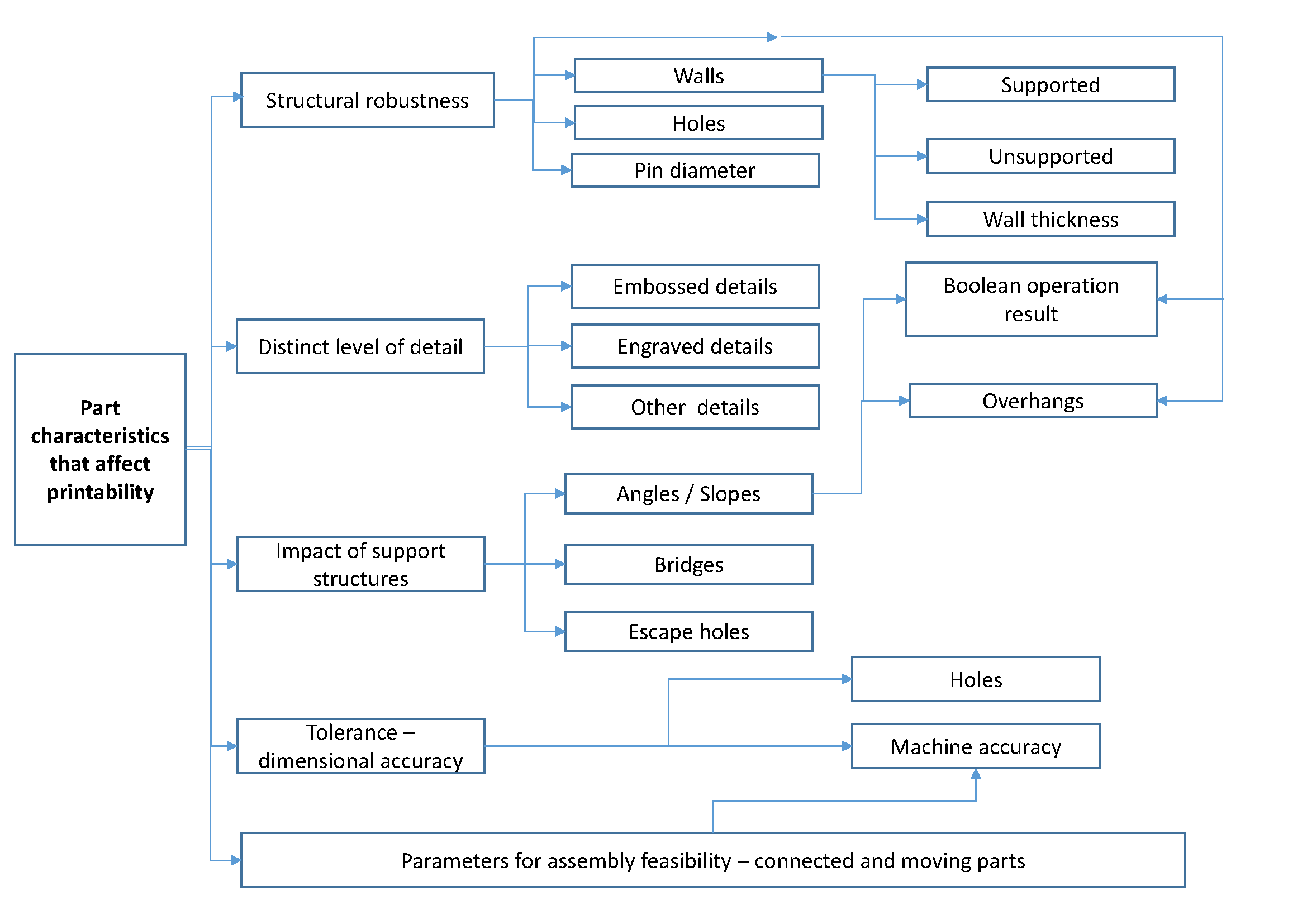}
\caption{Part characteristics (Figure courtesy of \citep{Fudos2020}).}
\label{fig:part_char}
\end{figure*}

The part characteristic probability function, $P_{F}(i,d,T,A)$, is a more decisive parameter in whether a model will be printed robustly, as it computes the probability of structural failure of a specific part characteristic (e.g hole, pin, etc), under given constraints. The $P_F$ expresses the probability of a part failing to be correctly printed (fabricated). The corresponding equation is improved by adding a new parameter $\epsilon$ expressing the mean absolute value, predicted by the ANN, of the dimensional error for a specific characteristic $i$ acquired by the error prediction method of Section \ref{sec:prediction}. This parameter is of the utmost importance because borderline dimensional values may affect the robustness of the corresponding part. Thus, for each part characteristic $i$ (Figure \ref{fig:part_char}) with a set of parameter dimensions $d(i)$, the part characteristic probability function $P_{F}(i,d,T,A)$ is expressed by Equation \ref{eq:pcpf}).

\begin{equation}
   P_{F}(i,d,T,A)=(1-\frac{1}{1+e^{(w_d(T,i)-(d-\epsilon))\,c}})\,s(A,i)
   \label{eq:pcpf}
\end{equation}

The threshold value $w_d(T,i)$ depends on the AM technology $T$ and the dimension type (length, diameter, etc) of the part characteristic ${i}$, and expresses the critical value for which the characteristic has a 50\% probability of printing successfully on the desired technology. When $d$ increases above the given threshold value $w_d(T,i)$, meaning that $P_F$ decreases, Equation \ref{eq:pcpf} is used to calculate the exact probability. 

The significance $0 < s(A,i) \leq 1$ expresses the impact of $i$ on the final printed part, in relation to an application $A$ and is set with respect to the violation of a design rule for a part characteristic with respect to functionality and robustness of the overall product. For example, the significance value of a small through hole is set close to $0$ for decorative artifacts and close to $1$ if it is used for ventilation or assembly of a mechanical part.

Table \ref{tab:critical} provides the critical values $w_d(T,i)$ for various part characteristics based on the design rules and constraints provided by several manufacturers and professionals for the three AM technologies: Fused deposition modeling (FDM), Binder Jetting (BJ) and Material Jetting (MJ) \citep{hubs_all,hydra_FDM,stratasys_polyjet,materialise}. In this study, the part characteristics chosen to develop this core version of the printability framework are related to structural robustness (supported/unsupported walls, pins, holes, overhangs, thin parts) and detail rendering (embossed/engraved details).     

\begin{table}[htbp]
\centering
\resizebox{9cm}{!}{
    \begin{tabular}{|p{5cm}|p{2.5cm}|p{3.0cm}|p{2.5cm}|}
         \hline
         \textbf{\large Characteristics}& \textbf{\large FDM} & \textbf{\large BJ} & \textbf{\large MJ}\\
         \hline
        Holes (Diameter) & 2.0mm & 1.5mm & 0.5mm\\
        Pins (Diameter) & 1.8mm & 2.0mm & 0.5mm\\
        Supported walls (Thickness) & 0.8mm & 2.0mm & 1.0mm\\
        Unsupported walls (Thickness) & 0.8mm & 3.0mm & 1.0mm\\
        Bridges (Length) & $\leq$ 10.0mm & - & - \\
        Thin parts (Thickness) & 2.0mm & 2.0mm & 0.5mm\\
        Overhangs (Angle) & $\leq$ \ang{45} & & \\
        Embossed details  & 0.6mm(width)   & 0.5mm (both) & 0.8mm(width) \\
        &                  2.0mm(height)   &              & 0.5mm(height)\\
        Engraved details  & 0.5mm(width)   & 0.5mm (both) & 0.5mm (both)\\
        &                  0.9mm(depth) & & \\
        \hline
    \end{tabular}
}
\caption{Critical values of $w_d(T,i)$ for part characteristics for each of three AM technologies based on the constructors (table by authors).}
\label{tab:critical}
\end{table}
 
The coefficient factor $c$ is obtained automatically for each dimension type $d$ of characteristic $i$. To find the optimal coefficient factor $c$, the following process was applied: based on the critical values $w_d(T,i)$ of characteristic $i$, two values were set as the lower and upper limits of the interval where $P_F$ tends to 1 and $P_F$ tends to 0 respectively. This interval is defined as $[d_{min},d_{max}]$ with the upper and lower limits set to $d_{max} =2 \, w_d(T,i)$ and $d_{min} = \frac{w_d(T,i)}{10 \, d_{max}}$. In this interval, the ideal setting is to have linear behavior for $P_F$. Since  a $C^1$ continuous function is preferred to satisfy this goal, the sigmoid-based function is used (Equation \ref{eq:pcpf}) by optimizing the following quantity for different $x$ values (where $x$ is the dimension of the part characteristic to be examined) with respect to coefficient factor $c$ (Equation \ref{eq:objective_1}):
\begin{equation}
\int_{d_{min}}^{d_{max}}((\frac{d_{max}-x}{d_{max}-d_{min}})-(1-\frac{1}{1+e^{(w_d(T,i)-x)\,c}}))^2\,dx
\label{eq:objective_1}
\end{equation}

The problem is attained by approximating the integral by a discrete sum and using a non linear optimization algorithm for computing coefficient factor $c$. Two optimization algorithms were used to test the accuracy and consistency of the results: Conjugate gradient \citep{congGrad} and L-BFGS \citep{lbfgsb}, both providing similar results. Furthermore, it was  determined analytically that Equation \ref{eq:objective_1} has a single global minimum. 
In cases where the probability $P_F$ of a part failing to print  decreases when dimension $d$ increases, Equation \ref{eq:pcpf} is used to determine the value of the $P_{F}$.
In other cases, for overhangs and bridges, where the probability $P_F$ for a part failing to print correctly increases when $d$ increases, Equation \ref{eq:pcpf_transf} is used to determine the $P_{F}$ value.

\begin{equation}
   P_{F}(i,d,T,A)=(\frac{1}{1+e^{(w_d(T,i)-(d-\epsilon))*c}})\,s(A,i))
   \label{eq:pcpf_transf}
\end{equation}

Subsequently, the objective function to be optimized is given by Equation \ref{eq:objective_2}.

\begin{equation}
\int_{d_{min}}^{d_{max}}((\frac{x-d_{min}}{d_{max}-d_{min}})-(\frac{1}{1+e^{(w_d(T,i)-x)\,c}}))^2\,dx
\label{eq:objective_2}
\end{equation}

\subsection{Verifying the part characteristic probability function for thin parts and pins}
\label{sec:evaluation pf}

 To substantiate the characteristic probability function $P_{F}$ of the printability framework, we conducted several experiments on how parameters, such as the thickness (for walls and thin parts in general) and diameter of pins, affect the maximum stresses applied to the model. The maximum stresses are responsible for the failure of the fabricated part. In this context,  failure is equivalent to collapsing. We also surveyed and thoroughly verified  information provided by manufacturers. The validation of the formulas of Section \ref{sec:printability parameters} is based on verifying two claims: 
 \begin{itemize}
     \item[C1.] The maximum stresses of the thin parts and pins depend linearly on the part thickness for typical length and width values.
     \item[C2.] The probability of failure of a fabricated part depends linearly on the maximum stress applied to the part. 
 \end{itemize}
 
 The verification of claims $C1$ and $C2$ is provided in Appendix A.


\subsection{Extending the repertoire of characteristic parts: the case of overhang structures}
\label{sec:overhangs}

Several researchers have considered the robustness of overhang structures by performing both theoretical and experimental studies (see e.g. \citep{Kim2018,Oropallo2016,Muthusamy2018,Jiang2018}).
To this end, we provide a method for determining the critical value $w_d(T,i)$  over which the overhang structure collapses. 

We conducted a study to examine, through the use of finite element analysis (FEA), the impact of several parameters on the behavior of the final printed overhang structure. For an overhang structure $O$, the length $L_O$, width $W_O$ and thickness $T_O$ at an angle $A_O$ are the factors modelled to experimentally determine the breaking point in terms of maximum von Mises equivalent stresses.

Based on this process, we can derive a critical maximum stress value $w_d(T,i)$ for the overhang structure with approximately $50\%$ probability of collapsing, which is then used in Equation \ref{eq:pcpf_transf} to determine the specific characteristic printability term. In the case of overhang structures, the dimensional error $\epsilon$ predicted by the neural network is only considered in setting the geometric dimensions that are given as input to the problem. 

\begin{figure}[htbp]
\centering
\includegraphics[scale=0.4]{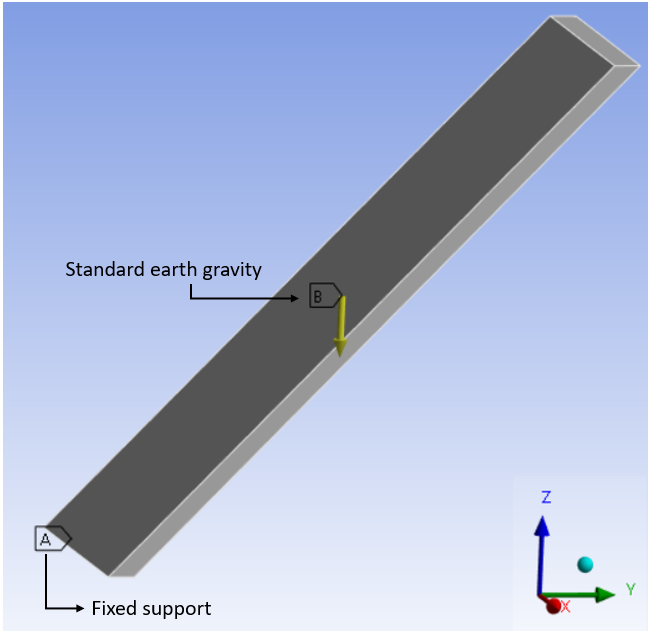}
\caption{Static structural analysis (figure by authors).}
\label{fig:static structural}
\end{figure}

For the analysis, we considered several different benchmarks of overhang geometries with $A_O$=$45\degree$, (based on the critical values provided by manufacturers in Table \ref{tab:critical}) for several lengths and thicknesses with a constant width value set to $W_O$=15mm. A static structural analysis was performed for each model under standard earth gravity ($9.8066\,m/s^2$), denoted as $B$ and applied in -Z axis with a fixed support on the models base, denoted as $A$ (Figure \ref{fig:static structural}). All models were meshed with linear tetrahedral elements. The 3D printer used for the experiments was ZPrinter 450 (BJ technology) and the material used was a composition of zp151 powder with z63 binder with a measured density equal to $1034\,kg/m^3$ without glue infiltration.

\begin{figure}[htbp]
\centering
    \includegraphics[scale=0.55]{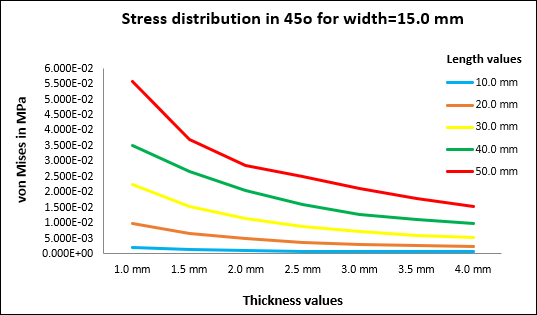}
    \caption{Maximum von Mises stress values for several cases of overhang geometries with $A_O=45\degree$ (figure by authors).}
\label{fig:stress graph}
\end{figure}  

\begin{figure*}
    \centering
    \includegraphics[scale=0.60]{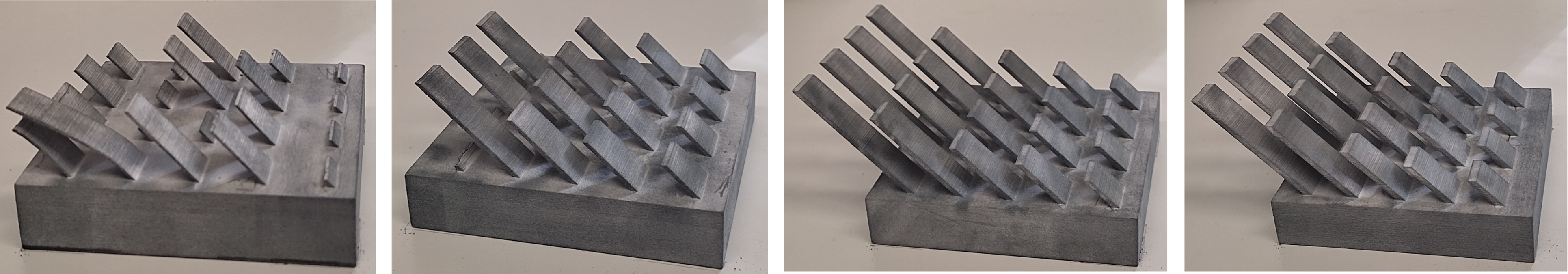}
\caption{Printed overhang geometries at $45\degree$ for $T_O=1.0mm, 1.5mm, 2.0mm, 2.5mm$ correspondingly (figure by authors).}
\label{fig:bench_overhangs}
\end{figure*}

The maximum stresses for all cases are shown in Figure \ref{fig:stress graph}. To determine the critical maximum stress value  which triggers fracture, all models were printed several times. Different benchmark models were designed, each of them capturing a sequence of repetitive patterns for different values of $T_O$ and $L_O$, for the same angle $A_O=45$\degree \, and $W_O=15\, mm$ (indicatively Figure \ref{fig:bench_overhangs}). Each of these benchmarks was printed four times meaning that each one of the overhang geometries with specific $T_O$ and $L_O$ was printed 16 times in total. The different dimensions used in the models were $T_O=1.0\,mm$ to $3.0\,mm$ with step=$0.5\,mm$ and $L_O=10.0\,mm$ to $50.0\,mm$ with step=$10.0\,mm$.

\begin{figure}[htbp]
\centering
\includegraphics[scale=0.55]{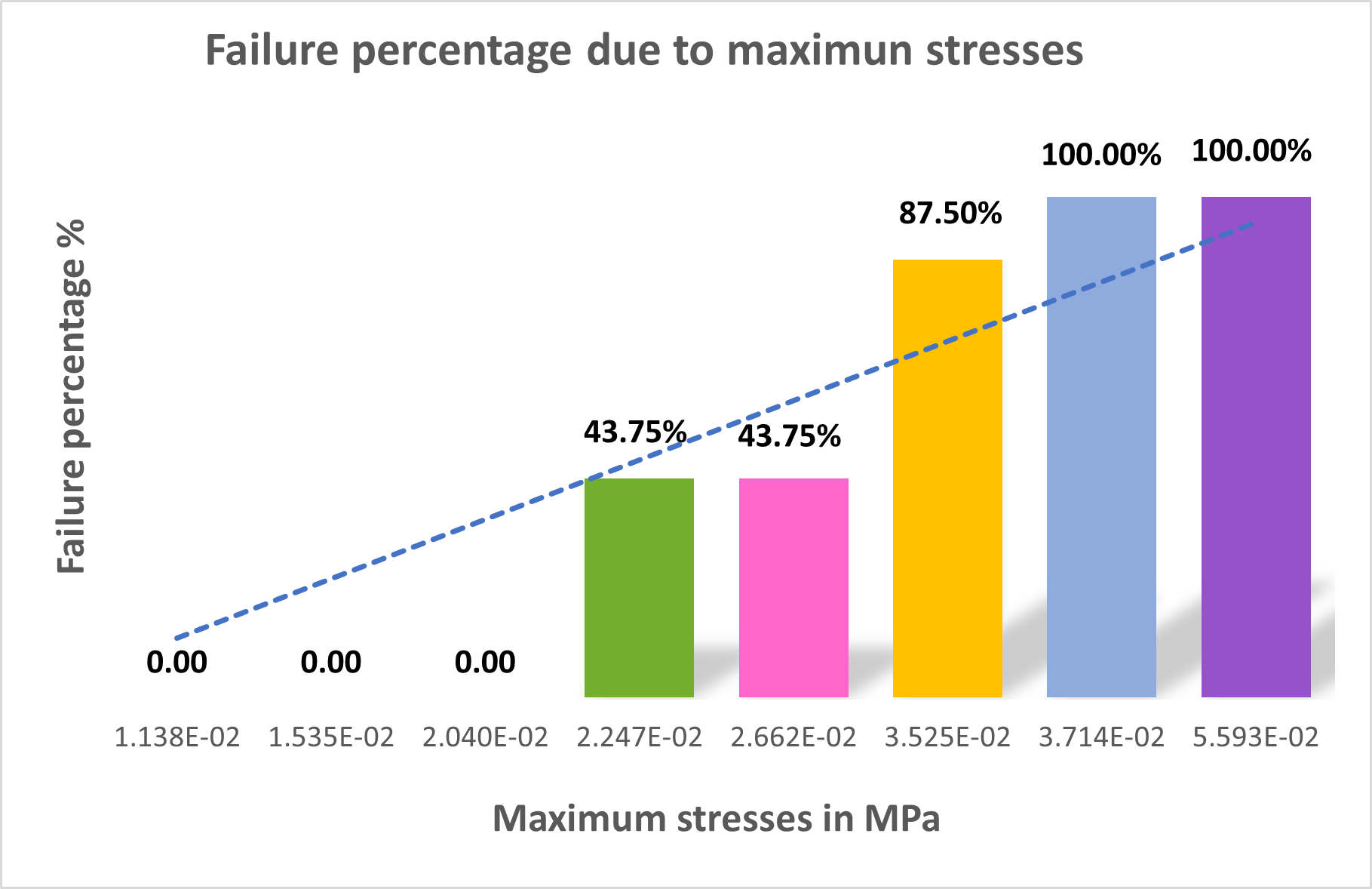}
\caption{Percentage of overhang geometry failures for different values of maximum stress values (figure by authors).}
\label{fig:failure graph}
\end{figure}

Based on these experimental results we have determined a rough estimation of the relationship between maximum stresses and printing failure percentage (see Figure \ref{fig:failure graph}). The critical maximum stress with 50\% probability of successful print was found to be: $w_d(T,i) \approx 2.919$E$-02\,MPa$.

\subsection{Printability computation tool}

\begin{figure*}
 \centering
 \includegraphics[width=0.7\linewidth]{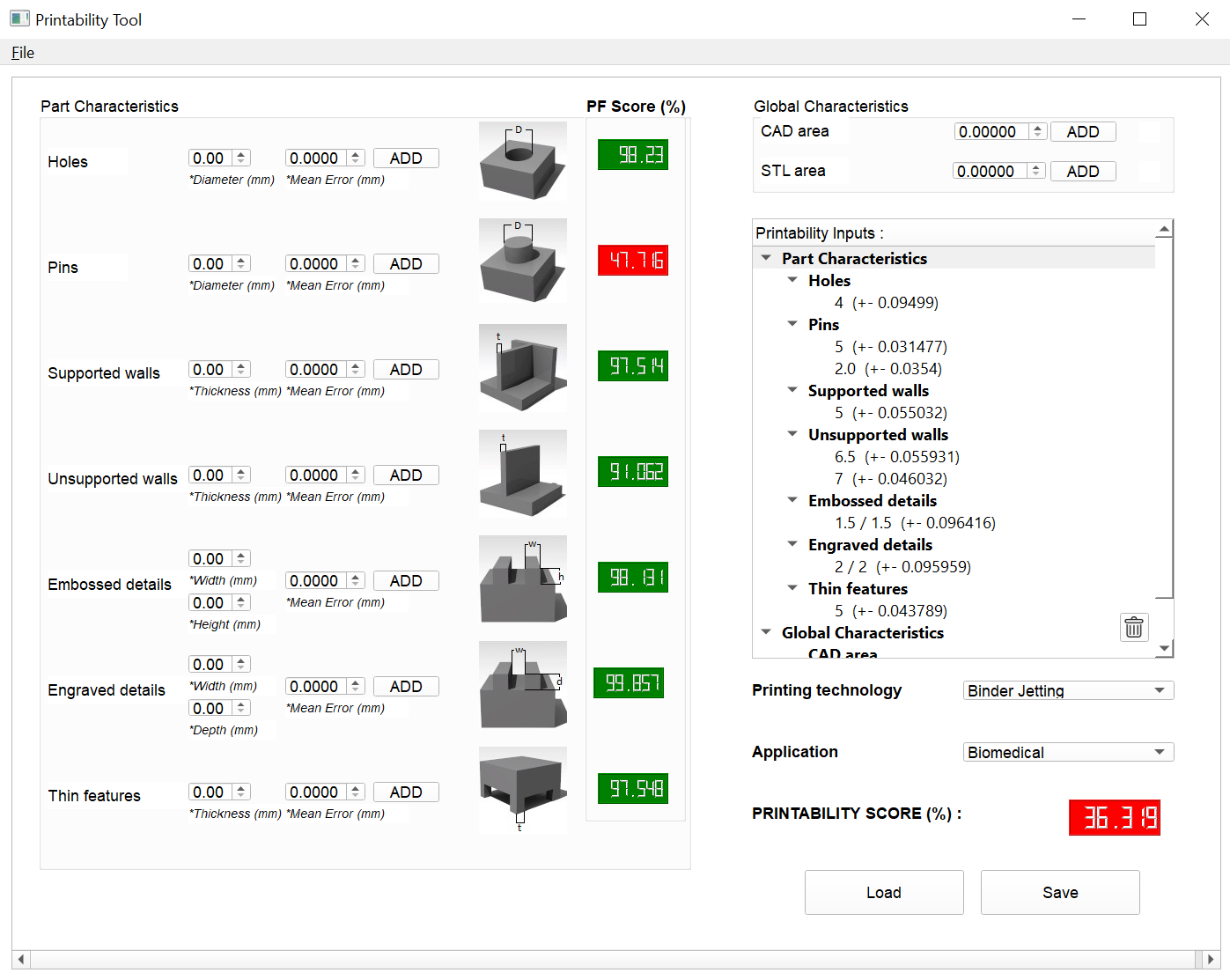}
 \caption{Overall printability score computation tool (figure by authors).}
 \label{fig:print_tool}
\end{figure*}

A user-friendly interactive tool that determines the printability score has been developed\footnote{Source code: \url{https://github.com/spirosmos/PrintabilityTool}}. A screenshot of the graphical user interface is shown in Figure \ref{fig:print_tool}. The user selects one of the three AM technologies and one of the three application domains. The user also specifies the dimensions of the critical geometric features of the mesh. The system supports many instances for each part characteristic category and an example of the required dimensions is provided next to each category. The CAD model and stl mesh surface areas are also provided as inputs for the computation of the global probability function (Equation \ref{eq:global}). All data are displayed in a tree structure from which the user can interactively remove any of them. Part characteristic probability for a part to be printed without failure $1-P_F$ (Equations \ref{eq:pcpf} and \ref{eq:pcpf_transf}) is displayed next to each part characteristic category whereas the final printability score is displayed at the bottom right corner of the window. The tool can store a specific configuration provided by the user as a .json file. Finally, there is a warning highlighting the printability score by red background color when the probability of successful printing is less than 50\%. 

\section{Experimental validation of the proposed methodology}

To evaluate and validate the proposed methodology several experiments were performed using various models. The models used for the experiments in this section are: a) ``Woman of Pindos", b) ``Dodone Eagle", c) ``Terracotta Warrior" (Table \ref{tab:free-form models}) and d) 3 benchmark models with the same set of part characteristics but of different dimensions (Tables \ref{tab:benchmarks-CAD}, \ref{tab:benchmark 1-printability}, \ref{tab:benchmark 2-printability}, \ref{tab:benchmark 3-printability}). All models were printed using ZPrinter 450. Each test model was printed several times, with a layer thickness of $0.102\,mm$. 

\begin{table}[htpp]
\centering
\resizebox{9cm}{!}{
\begin{tabular}{lll}
\rowcolor[HTML]{9698ED} 
\multicolumn{3}{c}{\cellcolor[HTML]{9698ED}\textbf{\Large{Free - Form Models}}} \\
\rowcolor[HTML]{DAE8FC} 
\multicolumn{1}{c}{\cellcolor[HTML]{DAE8FC}\textbf{\large{Woman of Pindos}}} & \multicolumn{1}{c}{\cellcolor[HTML]{DAE8FC}\textbf{\large{Dodone Eagle}}} & \multicolumn{1}{c}{\cellcolor[HTML]{DAE8FC}\textbf{\large{Terracotta Warrior}}} \\
\includegraphics[scale=0.35]{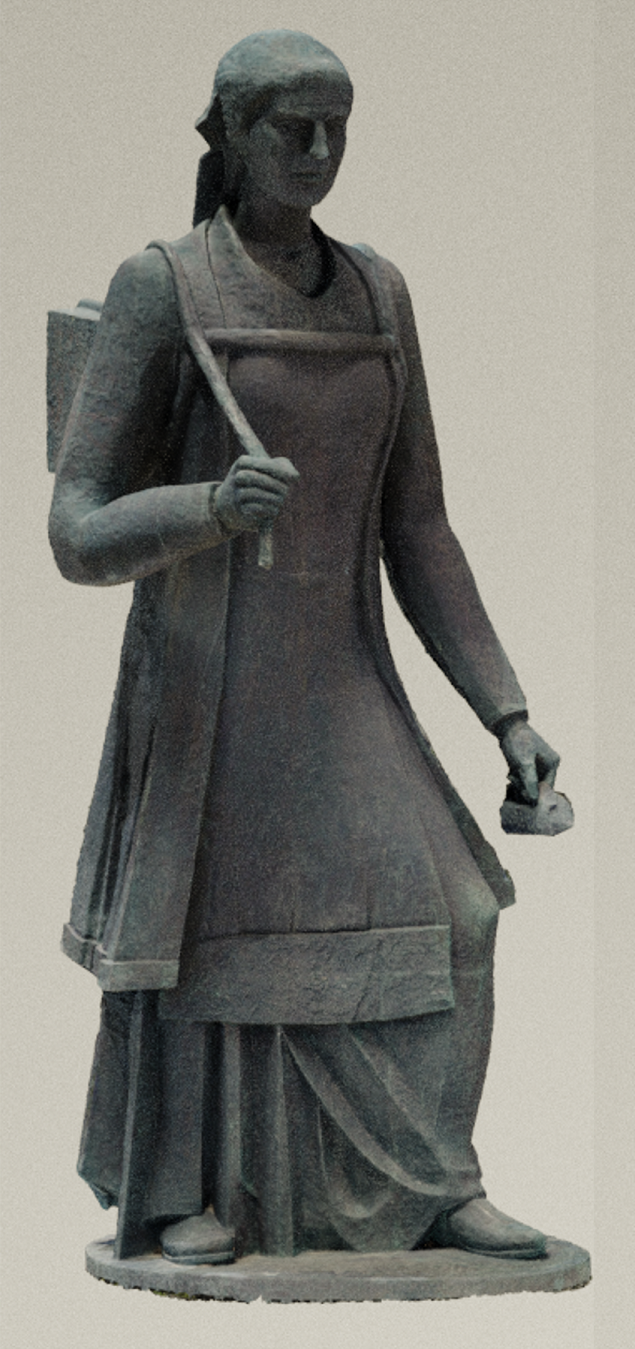} &
\includegraphics[scale=0.35]{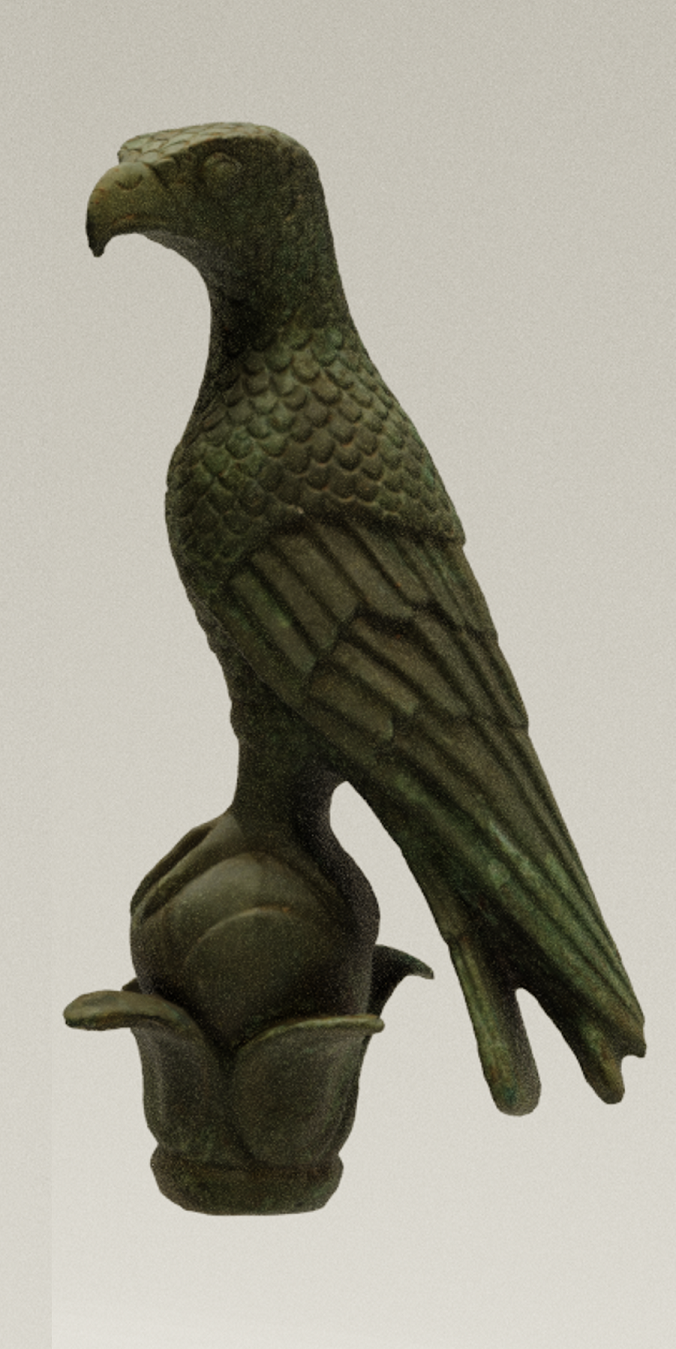} & 
\includegraphics[scale=0.35]{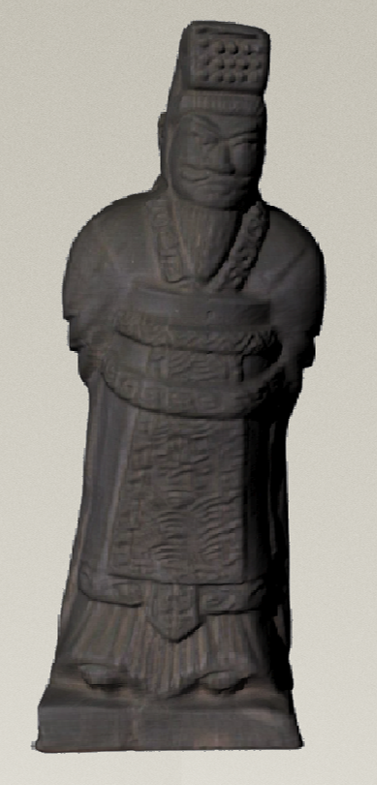}  \\
Mesh: $239028$ triangles  & Mesh: $87164$ triangles  & Mesh: $215600$ triangles    \\
Area: $0.011m^2$  & Area: $0.0038m^2$  & Area: $0.011m^2$  \\
Application: Art  & Application: Art  & Application: Art  \\
Reconstruction: Photogrammetry  & Reconstruction: Laser scanner & Reconstruction: Laser scanner  \\
Part characteristics: thin strap  & Part characteristics: None  & Part characteristics: None 
\end{tabular}
}
\caption{Free-form models (table by authors).}
\label{tab:free-form models}
\end{table}

\begin{table}[h!]
\centering
\resizebox{9cm}{!}{
\begin{tabular}{lll}
\rowcolor[HTML]{9698ED} 
\cellcolor[HTML]{9698ED}\textbf{\LARGE{Benchmark 1}} & \textbf{\LARGE{Benchmark 2}} & \textbf{\LARGE{Benchmark 3}} \\
\includegraphics[scale=0.35]{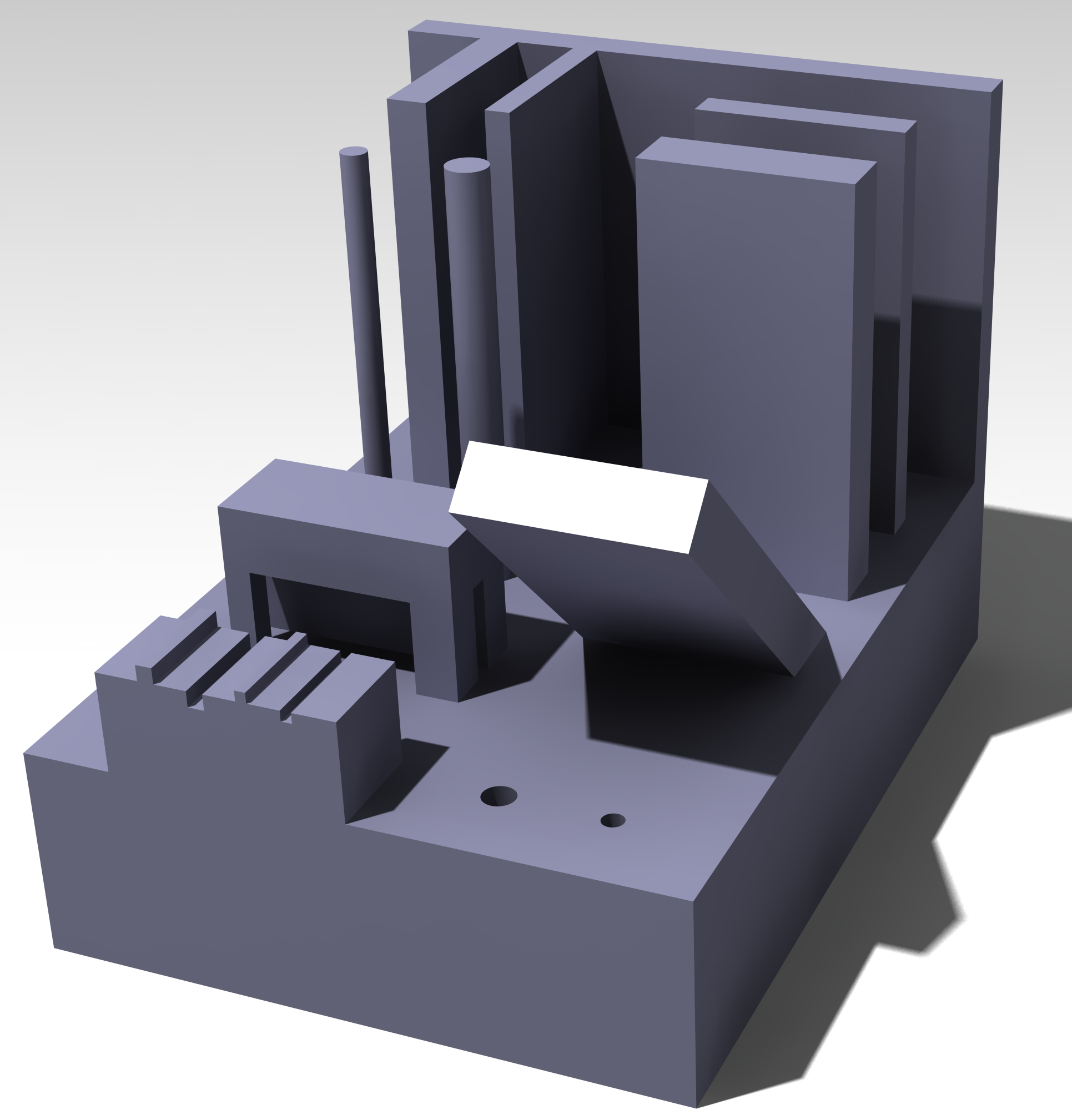}  &  \includegraphics[scale=0.35]{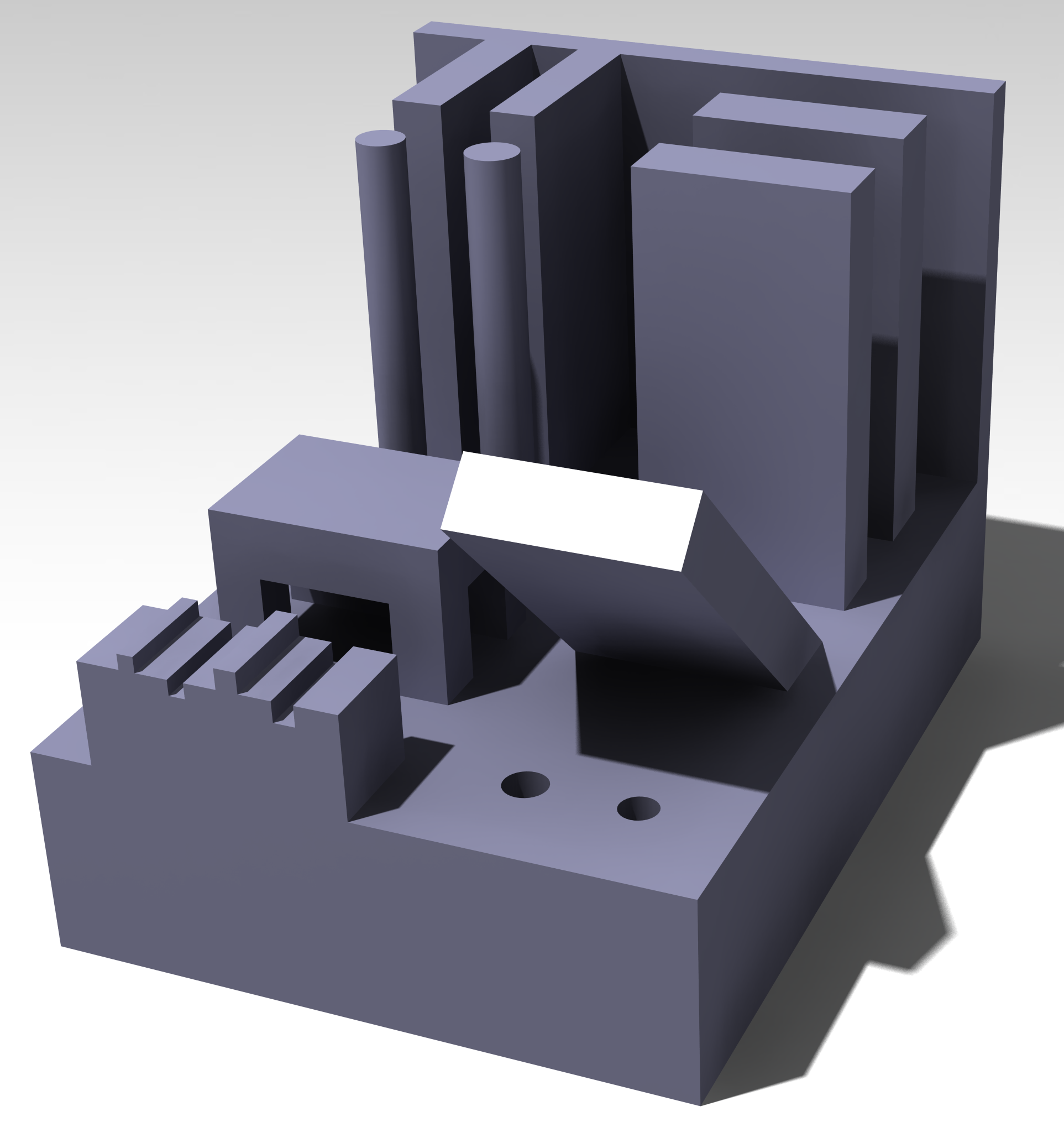}  &  \includegraphics[scale=0.35]{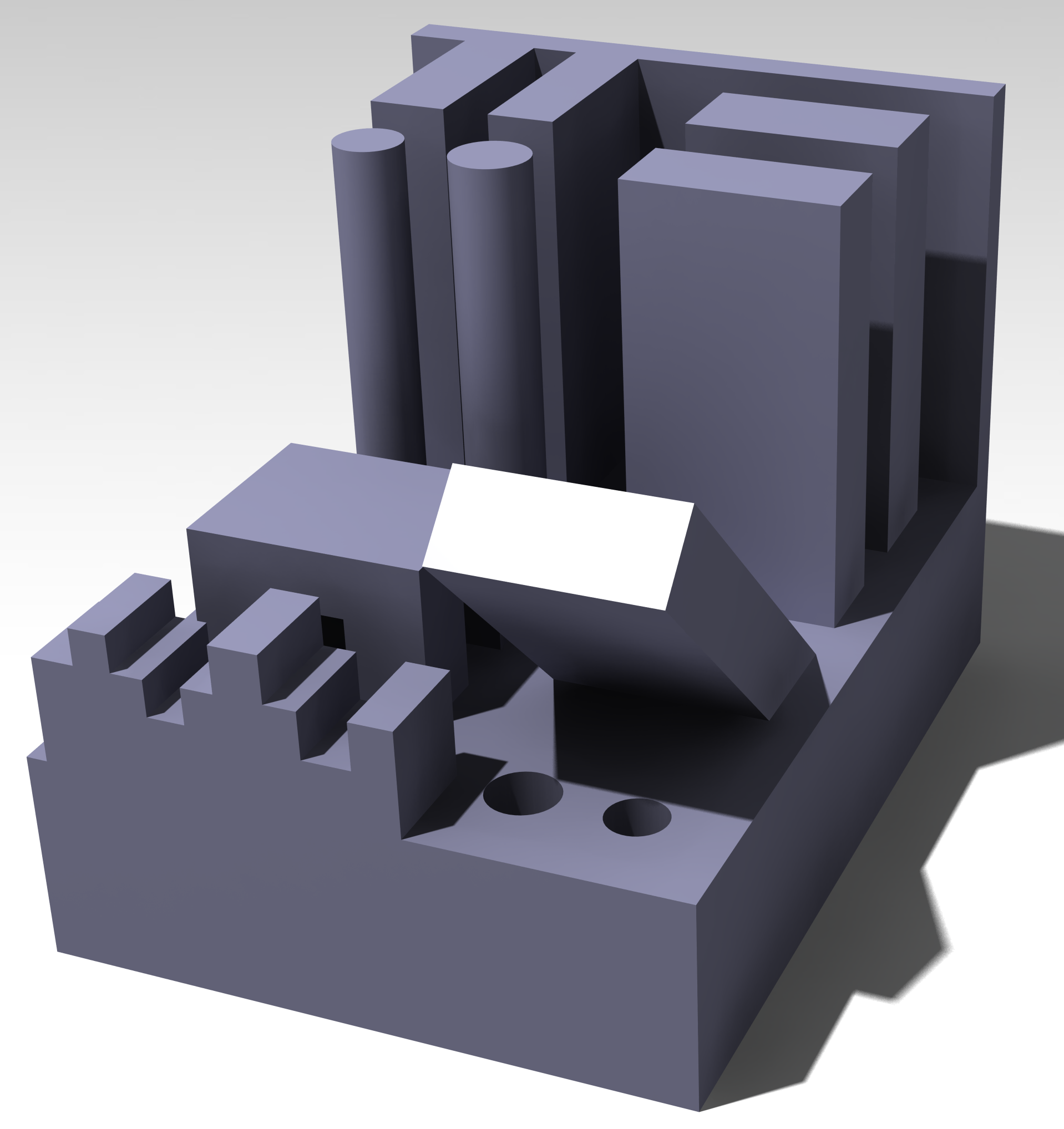}  \\
\Large{Mesh: $67654$ triangles}  &   \Large{Mesh: $69992$ triangles}  &   \Large{Mesh: $78470$ triangles}
\end{tabular}
}
\caption{CAD benchmark models of different dimensions (table by authors).}
\label{tab:benchmarks-CAD}
\end{table}

\subsection{Experimental evaluation of the reconstruction process}

\begin{figure}
\centering
  \begin{subfigure}[c]{0.4\columnwidth}
  \centering
    \includegraphics[width= 35mm]{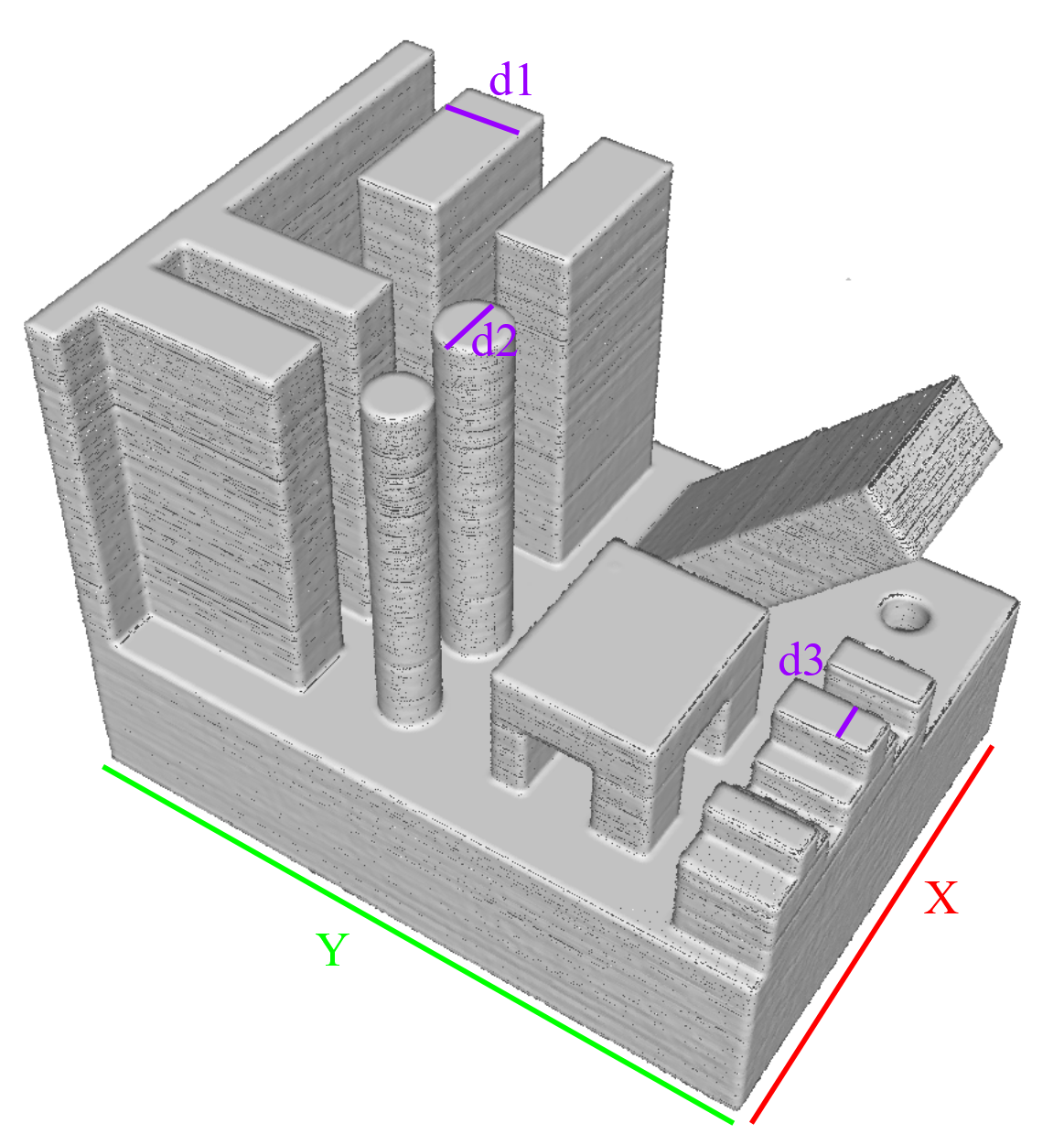}
    \caption{Our method.}
    \label{fig:scanner}
  \end{subfigure}
  \begin{subfigure}[c]{0.4\columnwidth}
    \includegraphics[width=35mm,trim={0mm 0 0mm 0mm},clip]{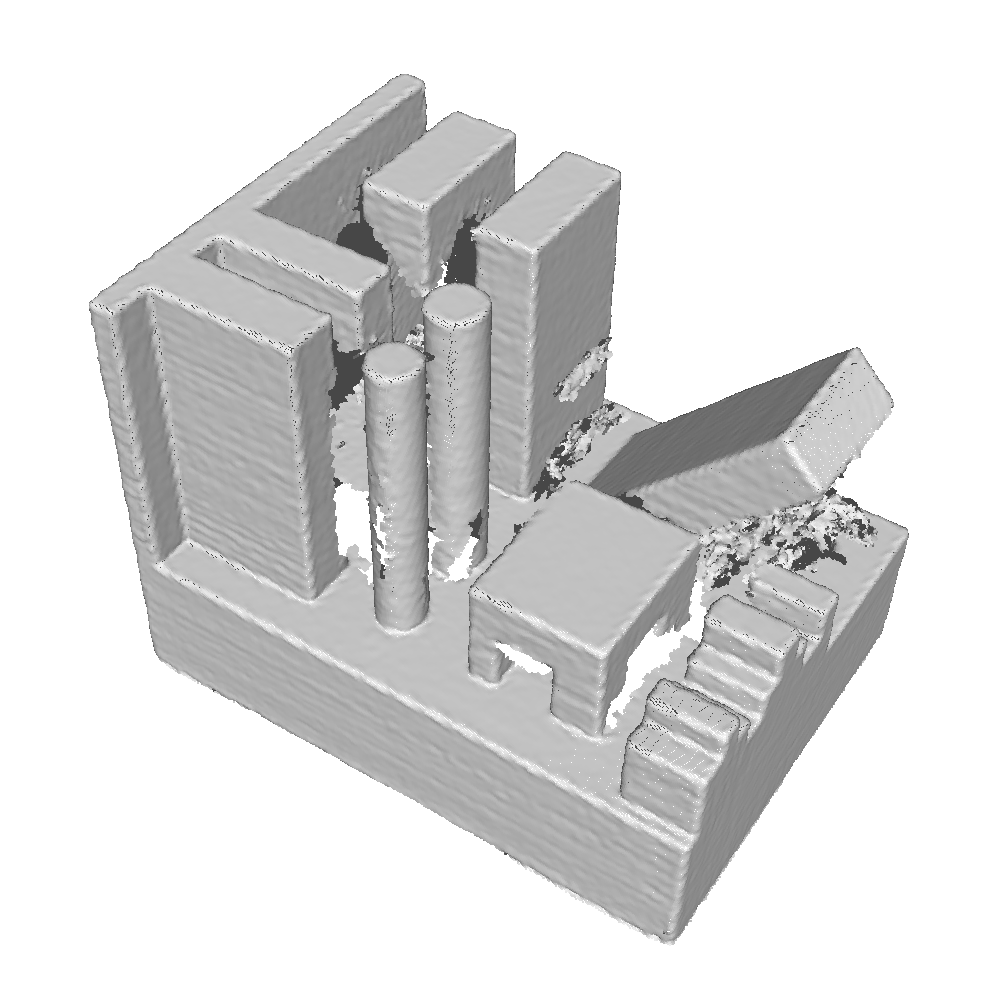}
    \caption{Einscan Pro+}
    \label{fig:reconstr}
  \end{subfigure}
\caption{Reconstructed point clouds (figure by authors).}
\label{fig:scannerRes}
\end{figure}

To test the reconstruction process presented in Section \ref{sec:recon} and evaluate its accuracy, we used Benchmark 3. We compare the point cloud derived from the reconstruction process of Section \ref{sec:recon} using a DJI Osmo Action Camera, the respective point cloud captured by a 3D laser scanner (Einscan Pro+ with Einscan HD Prime pack) and the actual fabricated object. In contrast to our method, the 3D scanner could not capture all points corresponding to the surface of the model (Figure \ref{fig:scannerRes}), producing a misleading dimensional error.

To evaluate the accuracy of the method, we focused on and compared the $x$ and $y$ dimensions of the bounding box, the thickness of a pin, an unsupported wall and an embossed wall from the fabricated model and both point clouds. A digital caliper was used for all measurements of the fabricated model. In Table \ref{table:measurements} we present the points of comparison used for the evaluation against the dimensions measured and the disparities in these measurements related to the CAD model. We observed that the values were similar in the cases of the pin and the unsupported wall for the 3D scanner and the reconstruction process. However, the rest of the results are in accordance, indicating that the reconstruction process is more reliable than that of the 3D scanner. The variations in the results can be attributed to the variable nature of factors such as perspective and lighting during the scanning process. Conversely, in the case of reconstruction, these factors are maintained constant, as the camera remains stationary and the lighting is stable.

\begin{table*}[]
\centering
\resizebox{0.85\textwidth}{!}{
\begin{tabular}{c|c|c|c|c||c|c|c}
\hline
                 & CAD   & Fabricated & \textbf{Reconstruction} & 3D Scanner & Disparities (Fabricated) & \textbf{Disparities (Reconstruction)} & Disparities (3D Scanner) \\ \hline
Y                & 60.00 & 60.02      &   \textbf{60.15}    & 60.50          & 0.02             &   \textbf{0.15}            & 0.50                  \\ \hline
X                & 80.00 & 79.91      &   \textbf{80.14}    & 80.36          & -0.09            &   \textbf{0.14}            & 0.36                 \\ \hline
d1 (Unsupported) & 9.50  & 9.54       &   \textbf{9.58}     & 9.58           & 0.04             &   \textbf{0.08}            & 0.08                 \\ \hline
d2 (Pin)         & 7.50  & 7.46       &   \textbf{7.54}     & 7.47           & -0.04            &   \textbf{0.04}            & -0.04                 \\ \hline
d3 (Embossed)    & 4.50  & 4.60       &  \textbf{4.62}      &  4.90          & 0.10             &   \textbf{0.12}            & 0.40                 \\ \hline
\end{tabular}
}
\caption{Evaluation of dimensional accuracy (in mm) (table by authors).}
\label{table:measurements}
\end{table*}

\subsection{Experimental evaluation of the error prediction method}

\begin{figure}[htbp]
 \centering
 \includegraphics[width=0.9\linewidth]{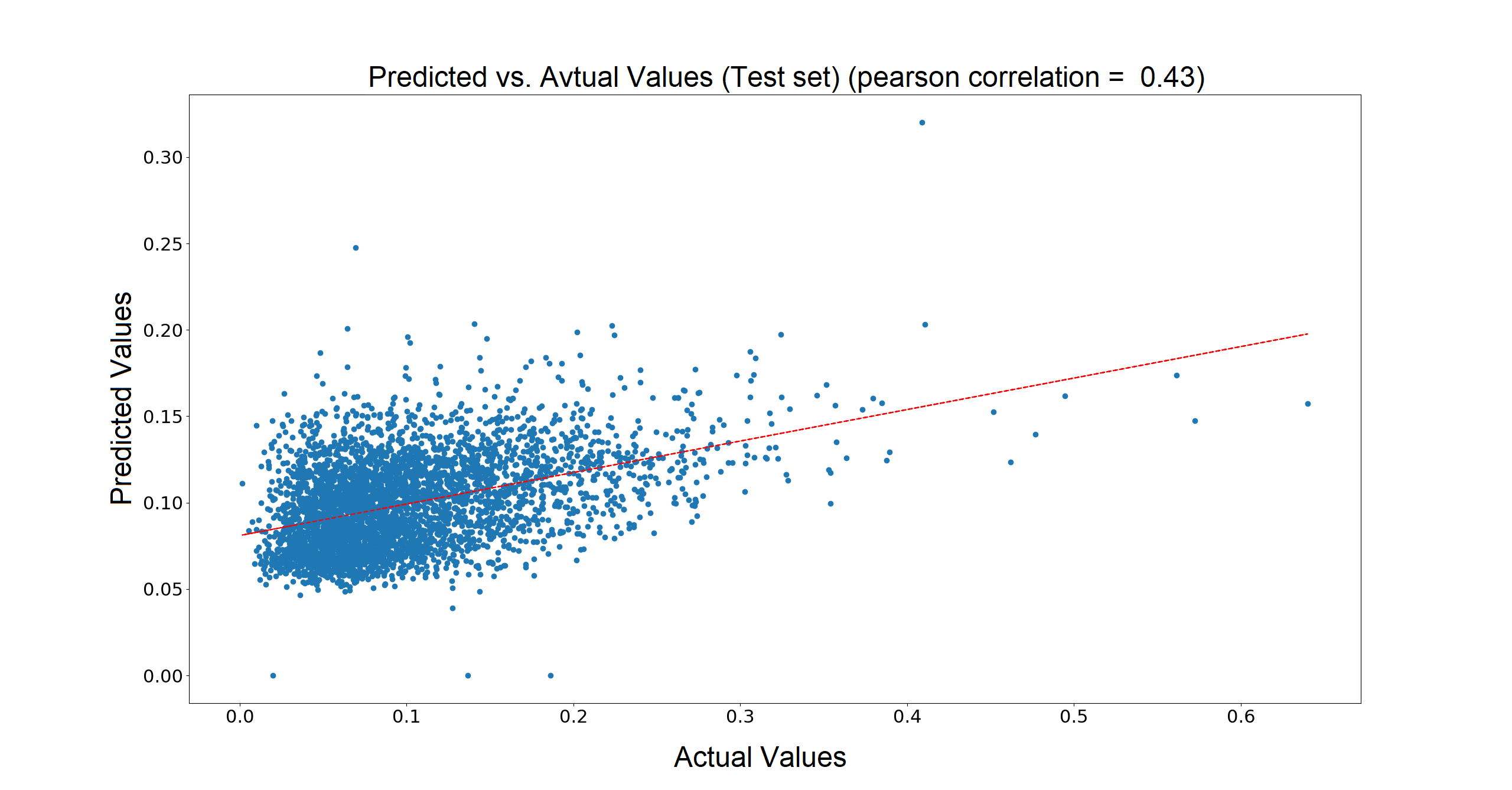}
 \caption{Pearson correlation between the actual and the predicted values (figure by authors).}
 \label{fig:pearson}
\end{figure}

To evaluate the predictions of the neural network we extracted a percentage of the data and used it as the test set. We cross validate the actual and predicted distance errors and measure the Pearson coefficient which demonstrates the linear correlation between them as depicted in Figure \ref{fig:pearson}. 

\begin{figure}[htbp]
 \centering
 \includegraphics[width=0.8\linewidth]{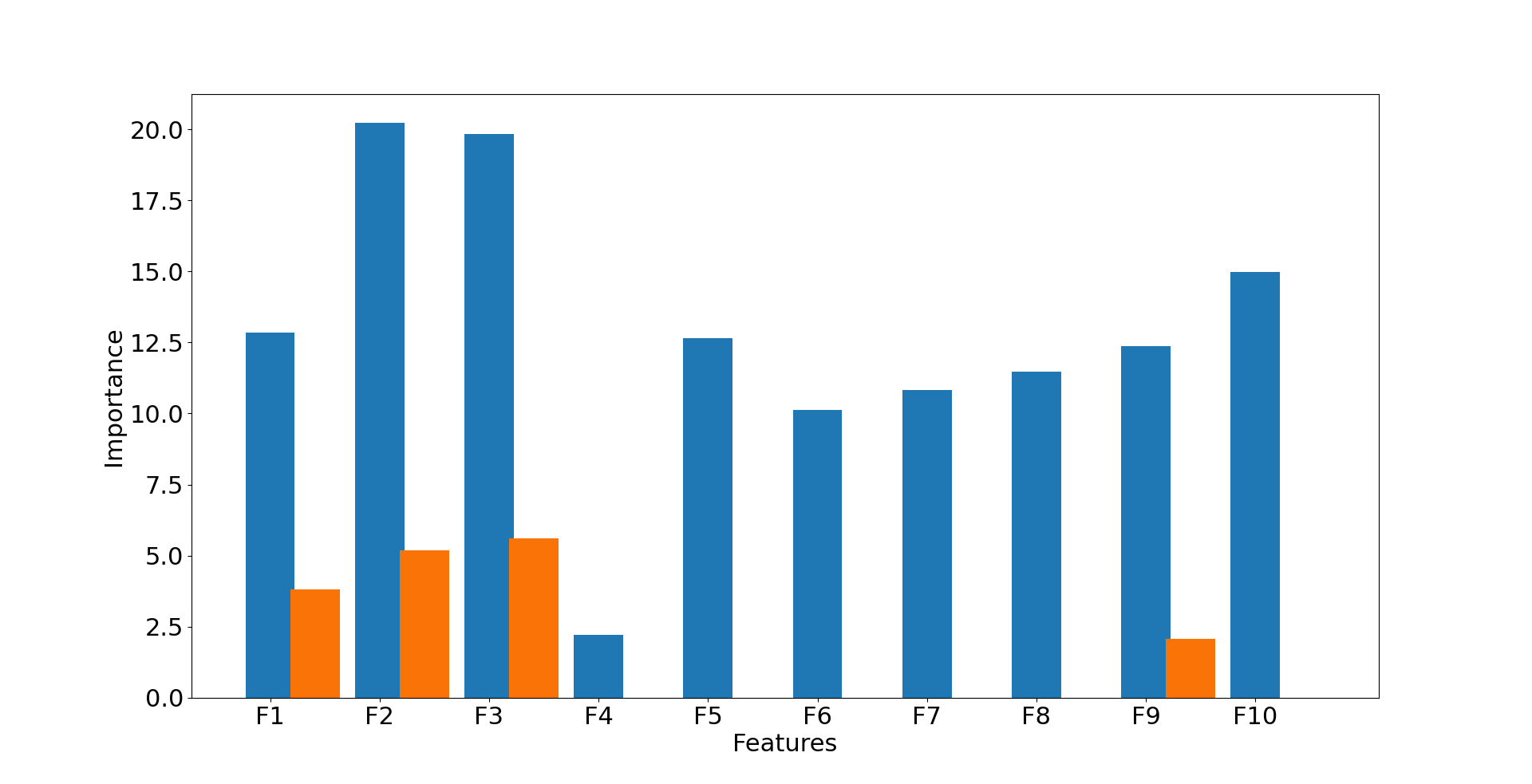}
 \caption{Feature importance: (blue) our features and (orange) similar predictors of \cite{DeckerHuang2019} (figure by authors).}
  \label{fig:importance}
\end{figure}

To compare our work with \cite{DeckerHuang2019}, we determined the importance of each feature for our prediction scheme by permuting the order of the values and observing how network loss is affected. Specifically, feature permutation was repeated 30 times. The difference between the actual and predicted MSE was calculated for both the shuffled and the correct feature sequence. Finally,  importance is given by normalizing the mean value of those differences by their standard deviation (the specifics of this process are provided in Section 4.2 of \cite{DeckerHuang2019}). Figure \ref{fig:importance} shows the results of this measure. From the results, it is clear all the features play a significant part in error prediction with importance values in $[2.0, 20.0]$ except for the Gaussian curvature (probably due to the high correlation with the mean curvature). In \cite{DeckerHuang2019} the importance of the predictors lies in $[1.7, 5.7]$, where the coordinate z yields the maximum importance whereas the {\em elevation angles} yield the minimum ($\approx 1.7$). Corresponding values for the importance of the features of our method and the importance of the predictors of \cite{DeckerHuang2019} exhibit a consistent behavior. 

\begin{figure}[htbp]
 \centering
 \begin{subfigure}[c]{0.45\columnwidth}
  \centering
    \includegraphics[width=0.4\linewidth]{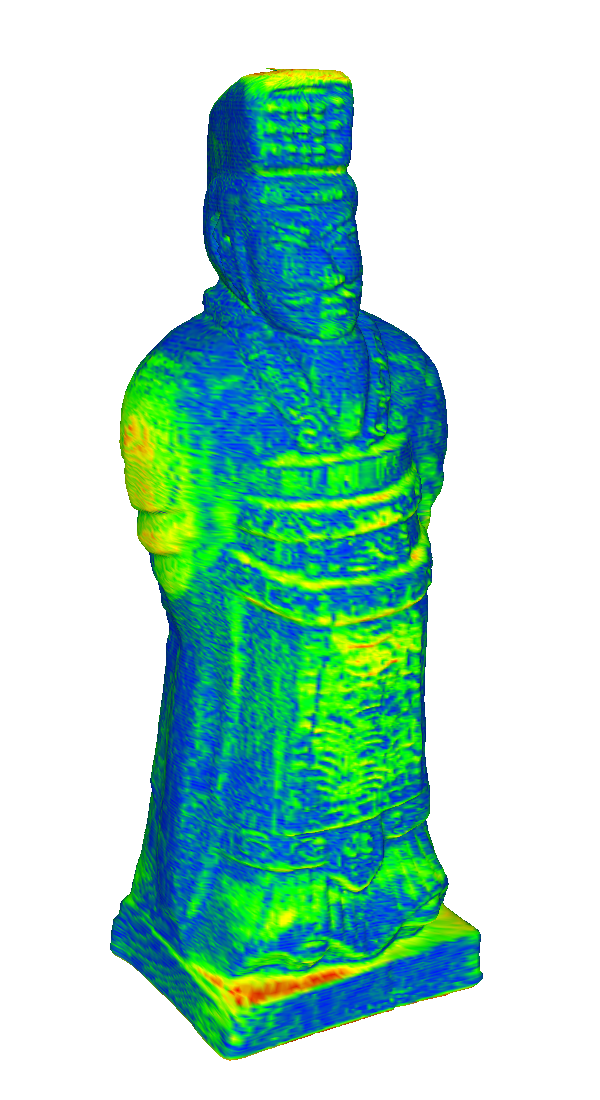}
    \includegraphics[width=0.4\linewidth]{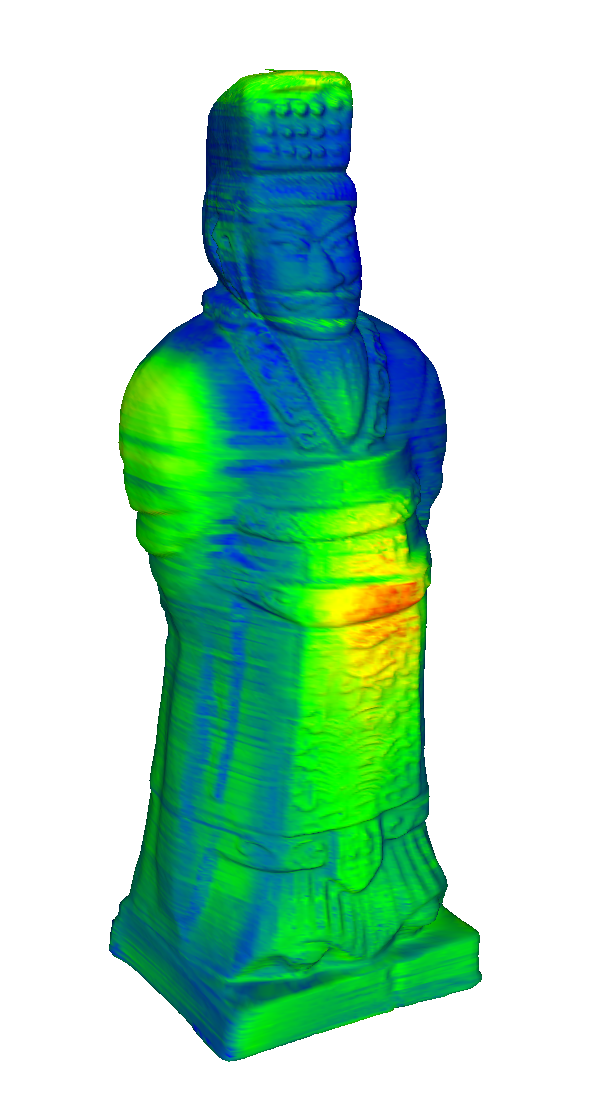}
    \caption{ Actual error vs Predicted error}
    \label{fig:Warrior_colors}
  \end{subfigure}
  \begin{subfigure}[c]{0.45\columnwidth}
    \includegraphics[width=0.44\linewidth]{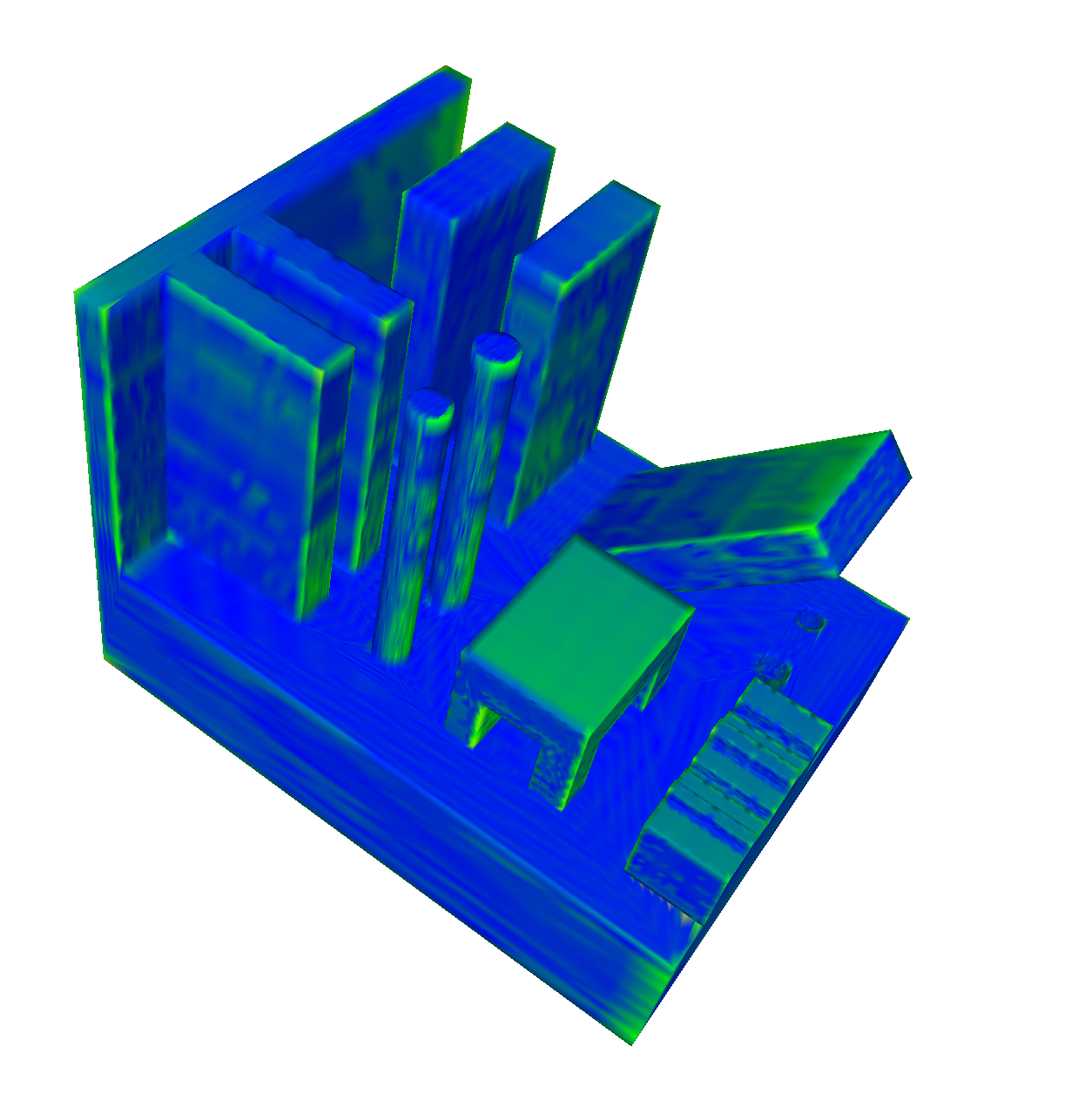}
    \includegraphics[width=0.44\linewidth]{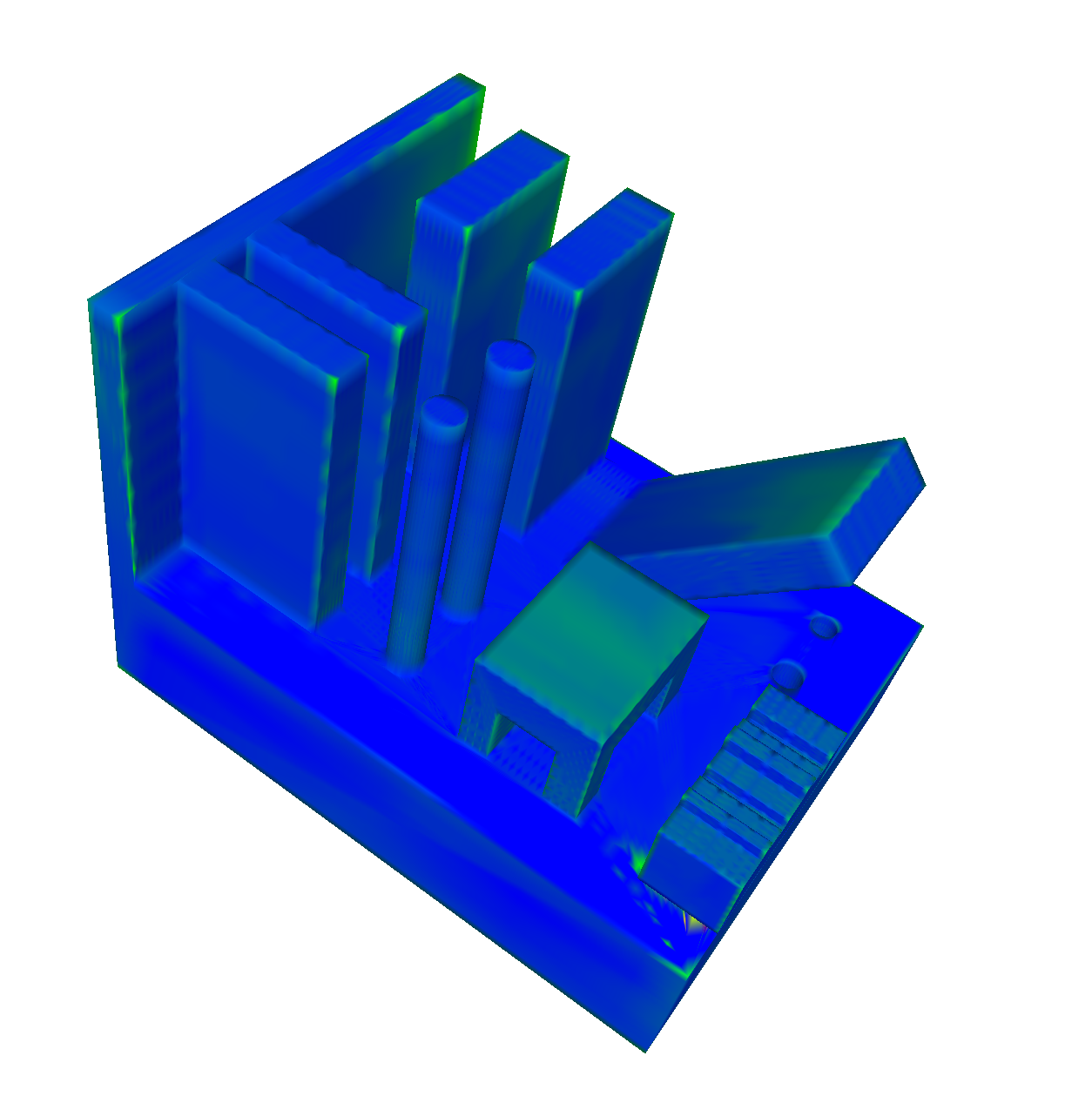}
    \caption{ Actual error vs Predicted error}
    \label{fig:Case3_colors}
  \end{subfigure}
 \caption{Coloring the error per vertex (actual and predicted) (figure by authors).}
  \label{fig:errorColoring}
\end{figure}

In addition, we tested the error prediction of the ANN by providing geometries whose actual dimensional errors were calculated using the C2C method and were not part of the training set. We computed the mean absolute dimensional error (MAE) and standard deviation (STD) of the predicted and actual values for each model (Table \ref{table:ann_pred1}). Our findings demonstrate that the ANN is sufficiently accurate based on both the MAE values and error distribution on the surface of the object (Figure \ref{fig:errorColoring}).

\begin{table}[htbp]
\centering
\resizebox{9cm}{!}{
\begin{tabular}{c|c|c||c|cl}
\cline{1-5}
\multicolumn{1}{l|}{} & \multicolumn{2}{c||}{\textbf{Actual Error (mm)}} & \multicolumn{2}{c}{\textbf{Predicted Error (mm)}} &              \\ \cline{1-5}
Models                 & MAE                   & STD                  & MAE                     & STD                    &               \\ \cline{1-5}

Benchmark 1           & 0.13521               & 0.10566              & 0.13076                 & 0.12027                &        \\
Benchmark 2           & 0.09918               & 0.07669              & 0.10220                 & 0.12326                &        \\
Benchmark 3           & 0.10565               & 0.11059              & 0.11897                 & 0.22477               &        \\ 
Terracotta warrior    & 0.08978               & 0.06283              & 0.08645                 & 0.06019                &       
\\\cline{1-5}
\end{tabular}
}
\caption{Actual and predicted MAE and STD (table by authors).}
\label{table:ann_pred1}
\end{table}

\subsection{Evaluating the printability framework}

\begin{figure}[h!]
\centering
\includegraphics[scale=0.35]{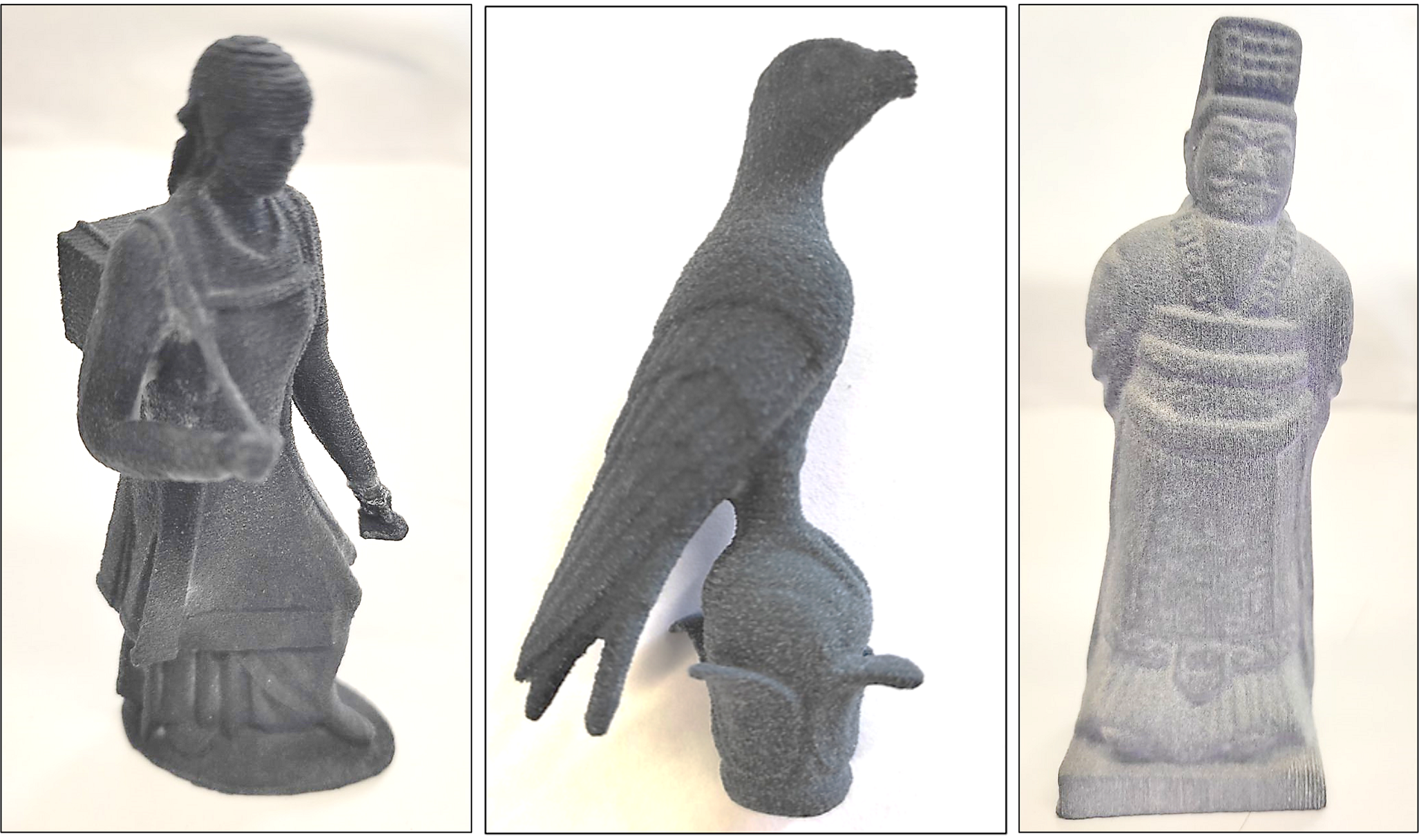}
\caption{From right to left:``Woman of Pindos", ``Dodonian Eagle", ''Terracotta warrior" fabricated parts (figure by authors).}
\label{fig:fabricated free form}
\end{figure}

To evaluate the validity of our printability estimation framework, all models were printed several times to evaluate the percentages of successful printing (overall printability score).
The ''Woman of Pindos" model has a specific part characteristic that directly affects its printability: the thin strap of the knapsack with a $1.5\,mm$ diameter. The critical value $w_d(T,i)$ for thin features in BJ technology is $2\,mm$, as listed in Table \ref{tab:critical}. The knapsack strap is an important element of the statue from an artistic standpoint, therefore, the application sensitivity parameter $s$ was set to 1. Because this model was reconstructed by laser scanning, there was no original CAD model and  $Q_{S_{CM}}=1$. The parameters for the printability calculation are listed in Table \ref{tab:my-table}. The probability of successfully printing the thin strap was $30.00\%$, a result that was verified after printing the models, because in only two out of the  six prints the strap was printed successfully. The overall printability score (see Equation \ref{eq:overall_score}) was estimated to be $27.38\%$. 
Both the ''Dodone Eagle" and the ''Terracotta warrior" are models of particular interest due to the peculiarity of the surface texture and detail. Because they contain no other geometrical characteristics, the $P_{G}$ calculation is equivalent to the overall printability and the sensitivity parameter $s$ was set to 1.  Because both models are reconstructed by laser scanning, there was no original CAD model and  $Q_{S_{CM}}=1$.  The overall printability score for both models was $91.28\%$ as presented in Tables \ref{tab:printability eagle} and \ref{tab:printability terracotta} respectively. Both models were succesfully printed six times, with no significant problems. 
The three freeform fabricated models, without glue infiltration treatment, are depicted in Figure \ref{fig:fabricated free form}.

\begin{table}[]
\resizebox{9cm}{!}{
\begin{tabular}{|ccc|}
\hline
\multicolumn{3}{|c|}{\cellcolor[HTML]{CBCEFB}\textbf{Woman of Pindos}}                                                 \\ \hline
\multicolumn{1}{|c|}{Thin part(x)=1.5 mm} & \multicolumn{1}{c|}{$w_d(T,i)=2.0\,mm$}        & Predicted error=0.18 mm     \\ \hline
\multicolumn{1}{|c|}{$D_{ST}$(Accuracy)=0.3}  & \multicolumn{1}{c|}{$D_{ST}$(Surface Texture)=0.3} & $D_{ST}$(Abnormalities)=0.3      \\ \hline
\multicolumn{1}{|c|}{s(A,i)=1}           & \multicolumn{1}{c|}{k(x,A)=0.9}               & $Q_{S_{CM}}=1$                   \\ \hline
\multicolumn{1}{|c|}{(1-$P_F$ )=0.300}      & \multicolumn{1}{c|}{(1-$P_G$)=0.9128}            & Overall Printability=0.2738 \\ \hline
\multicolumn{3}{|c|}{\cellcolor[HTML]{C0C0C0}\textbf{OPS=27.38\%}}                                      \\ \hline
\end{tabular}
}
\caption{Overall printability score for ”Woman of Pindos” (table by authors).}
\label{tab:my-table}
\end{table}

\begin{table}[]
\centering
\resizebox{8.5cm}{!}{
\begin{tabular}{|ccc|}
\hline
\multicolumn{3}{|c|}{\cellcolor[HTML]{CBCEFB}\textbf{Dodone Eagle}}                                                                         \\ \hline
\multicolumn{1}{|c|}{$D_{ST}$(Accuracy)=0.3}        & \multicolumn{1}{c|}{k(x,A)=0.9} & (1-$P_G$)=0.9128                                              \\ \hline
\multicolumn{1}{|c|}{$D_{ST}$(Surface Texture)=0.3} &  & Overall Printability=0.91268                                \\ \hline
\multicolumn{1}{|c|}{$D_{ST}$(Abnormalities)=0.3}   &  \multicolumn{1}{c|}{$Q_{S_{CM}}=1$}  & \cellcolor[HTML]{C0C0C0}\textbf{OPS=91.28\%} \\ \hline
\end{tabular}
}
\caption{Overall printability score for "Dodone Eagle" (table by authors).}
\label{tab:printability eagle}
\end{table}
\begin{table}[]
\centering
\resizebox{9cm}{!}{
\begin{tabular}{|ccc|}
\hline
\multicolumn{3}{|c|}{\cellcolor[HTML]{CBCEFB}\textbf{Terracotta Warrior}}                                                                   \\ \hline
\multicolumn{1}{|c|}{$D_{ST}$(Accuracy)=0.3}        & \multicolumn{1}{c|}{k(x,A)=0.9} & (1-$P_G$)=0.9128                                              \\ \hline
\multicolumn{1}{|c|}{$D_{ST}$(Surface Texture)=0.3} &  & Overall Printability=0.91268                                \\ \hline
\multicolumn{1}{|c|}{$D_{ST}$(Abnormalities)=0.3}   &  \multicolumn{1}{c|}{$Q_{S_{CM}}=1$}  & \cellcolor[HTML]{C0C0C0}\textbf{OPS=91.28\%} \\ \hline
\end{tabular}
}
\caption{Overall printability score for "Terracotta warrior" (table by authors).}
\label{tab:printability terracotta}
\end{table}

Finally, in reference to the benchmark models, detailed information on achieving the overall printability score are presented in Tables \ref{tab:benchmark 1-printability}, \ref{tab:benchmark 2-printability} and \ref{tab:benchmark 3-printability}. Three different $k$ factors were used to test the effect of parameters on the overall printability score. The sensitivity factor $k=0.1$ leads to $(1-{P_G)}$=0.9900, $k=0.5$ to $(1-{P_G)}$=0.9508 and finally $k=0.9$ to $(1-{P_G)}$=$0.91288$ probabilities for successful printing.

\begin{table}[h]
\centering
\resizebox{8cm}{!}{
\begin{tabular}{cccc}
\rowcolor[HTML]{9698ED} 
\multicolumn{4}{c}{\cellcolor[HTML]{9698ED}\textbf{Benchmark 1}}                                                                                              \\
\rowcolor[HTML]{DAE8FC} 
\textbf{Part characteristics}                & \textbf{Dimensions} & \textbf{1-PF}     & \textbf{Equation}                             \\
\cellcolor[HTML]{EFEFEF}\textbf{Holes}       & D=3.0 mm               & 0.9117      &   \ref{eq:pcpf}                                            \\
\cellcolor[HTML]{EFEFEF}                     & D=2. mm               & 0.6604     & \multirow{-2}{*}{}                            \\
\cellcolor[HTML]{EFEFEF}\textbf{Pins}        & D=4.0 mm               & 0.9156      &  \ref{eq:pcpf}                                             \\
\cellcolor[HTML]{EFEFEF}                     & D=2.5 mm               & 0.6248      & \multirow{-2}{*}{}                            \\\cellcolor[HTML]{EFEFEF}\textbf{Unsupported} & t=6.0 mm               & 0.9189      &   \ref{eq:pcpf}                                           \\
\cellcolor[HTML]{EFEFEF}                     & t=3.5 mm               & 0.5878      & \multirow{-2}{*}{}                            \\
\cellcolor[HTML]{EFEFEF}\textbf{Supported}   & t=4.0 mm               & 0.9162      & \ref{eq:pcpf}                                              \\
\cellcolor[HTML]{EFEFEF}                     & t=2.5 mm               & 0.688      & \multirow{-2}{*}{}                            \\
\cellcolor[HTML]{EFEFEF}\textbf{Embossed}    & h=1.0 mm x w=1.0 mm       & 0.8988      &  \ref{eq:pcpf}                                             \\
\cellcolor[HTML]{EFEFEF}                     & h=1.5 mm x w=1.5 mm       & 0.9965      & \multirow{-2}{*}{}                            \\
\cellcolor[HTML]{EFEFEF}\textbf{Engraved}    & d=1.0 mm x w=1.0 mm       & 0.8992      &  \ref{eq:pcpf}                                             \\
\cellcolor[HTML]{EFEFEF}                     & d=1.5 mm x w=1.5 mm       & 0.9965      & \multirow{-2}{*}{}                            \\
\cellcolor[HTML]{EFEFEF}\textbf{Thin parts}  & t=4.0 mm               & 0.9135      & \ref{eq:pcpf}                                              \\
\cellcolor[HTML]{EFEFEF}                     & t=2.5 mm               & 0.6178      & \multirow{-2}{*}{}                            \\
\cellcolor[HTML]{EFEFEF}\textbf{Overhangs}   & s=$1.856E$+$04\,N/m^2$       & 0.6882      &   \ref{eq:pcpf_transf}                                             \\                                            &                     & \cellcolor[HTML]{E0C2CD}\textbf{$\prod (1-P_{F})$} & \cellcolor[HTML]{E0C2CD}\textbf{0.03341} \\
\rowcolor[HTML]{CBCEFB} 
\textbf{Sensitivity factor}                  & \textbf{1-PG}       & \textbf{P(M,T)}                          & \textbf{OPS (\%)}                   \\
for k=0.1                                    & 0.99004             & 0.03308                                  & \textbf{3.31\%}                               \\
for k=0.5                                    & 0.95089             & 0.03177                                  & \textbf{3.18\%}                               \\
for k=0.9                                    & 0.91288             & 0.03050                                  & \textbf{3.05\%}                              
\end{tabular}
}
\caption{Overall printability score calculation for Benchmark 1 (D:diameter, t:thickness, h:height, w:width, d:depth and s:stress) (table by authors).}
\label{tab:benchmark 1-printability}
\end{table}

\begin{table}[h]
\centering
\resizebox{8cm}{!}{
\begin{tabular}{cccc}
\rowcolor[HTML]{9698ED} 
\multicolumn{4}{c}{\cellcolor[HTML]{9698ED}\textbf{Benchmark 2}}                                                                                              \\
\rowcolor[HTML]{DAE8FC} 
\textbf{Part characteristics}                & \textbf{Dimensions} & \textbf{1-PF}     & \textbf{Equation}                             \\
\cellcolor[HTML]{EFEFEF}\textbf{Holes}       & D=3.5 mm               & 0.9602      &   \ref{eq:pcpf}                                            \\
\cellcolor[HTML]{EFEFEF}                     & D=4.0 mm               & 0.9822     & \multirow{-2}{*}{}                            \\
\cellcolor[HTML]{EFEFEF}\textbf{Pins}        & D=4.5 mm               & 0.9551      &  \ref{eq:pcpf}                                             \\
\cellcolor[HTML]{EFEFEF}                     & D=5. 0mm               & 0.9759      & \multirow{-2}{*}{}                            \\
\cellcolor[HTML]{EFEFEF}\textbf{Unsupported} & t=6.5 mm               & 0.9457      &   \ref{eq:pcpf}                                           \\
\cellcolor[HTML]{EFEFEF}                     & t=7.0 mm               & 0.9634      & \multirow{-2}{*}{}                            \\
\cellcolor[HTML]{EFEFEF}\textbf{Supported}   & t=5.0 mm               & 0.9752      & \ref{eq:pcpf}                                              \\
\cellcolor[HTML]{EFEFEF}                     & t=4.5 mm               & 0.9553      & \multirow{-2}{*}{}                            \\
\cellcolor[HTML]{EFEFEF}\textbf{Embossed}    & h=1.5 mm x w=1.5 mm       & 0.9960      &  \ref{eq:pcpf}                                             \\
\cellcolor[HTML]{EFEFEF}                     & h=2.0 mm x w=2.0 mm       & 0.9999      & \multirow{-2}{*}{}                            \\
\cellcolor[HTML]{EFEFEF}\textbf{Engraved}    & d=1.5 mm x w=1.5 mm       & 0.9960      &  \ref{eq:pcpf}                                             \\
\cellcolor[HTML]{EFEFEF}                     & d=2.0 mm x w=2.0 mm       & 0.9999      & \multirow{-2}{*}{}                            \\
\cellcolor[HTML]{EFEFEF}\textbf{Thin parts}  & t=4.5 mm               & 0.9527      & \ref{eq:pcpf}                                              \\
\cellcolor[HTML]{EFEFEF}                     & t=5.0 mm               & 0.9755      & \multirow{-2}{*}{}                            \\
\cellcolor[HTML]{EFEFEF}\textbf{Overhangs}   & s=$1.7825E$+$04\,N/m^2$       & 0.7036      &   \ref{eq:pcpf_transf}                                             \\                                            &                     & \cellcolor[HTML]{E0C2CD}\textbf{$\prod (1-P_{F})$} & \cellcolor[HTML]{E0C2CD}\textbf{0.4858} \\
\rowcolor[HTML]{CBCEFB} 
\textbf{Sensitivity factor}                  & \textbf{1-PG}       & \textbf{P(M,T)}                          & \textbf{OPS (\%)}                   \\
for k=0.1                                    & 0.99004             & 0.48104                                  & \textbf{48.30\%}                               \\
for k=0.5                                    & 0.95089             & 0.46202                                  & \textbf{46.39\%}                               \\
for k=0.9                                    & 0.91288             & 0.44355                                  & \textbf{44.54\%}                              
\end{tabular}
}
\caption{Overall printability score calculation for Benchmark 2 (D:diameter, t:thickness, h:height, w:width, d:depth and s:stress) (table by authors).}
\label{tab:benchmark 2-printability}
\end{table}

\begin{table}[h]
\centering
\resizebox{8cm}{!}{
\begin{tabular}{cccc}
\rowcolor[HTML]{9698ED} 
\multicolumn{4}{c}{\cellcolor[HTML]{9698ED}\textbf{Benchmark 3}}                                                                                              \\
\rowcolor[HTML]{DAE8FC} 
\textbf{Part characteristics}                & \textbf{Dimensions} & \textbf{1-PF}     & \textbf{Equation} \\
\cellcolor[HTML]{EFEFEF}\textbf{Holes}       & D=5.5 mm               & 0.9985      &   \ref{eq:pcpf}  \\
\cellcolor[HTML]{EFEFEF}                     & D=6.5 mm               & 0.9997     & \multirow{-2}{*}{}  \\
\cellcolor[HTML]{EFEFEF}\textbf{Pins}        & D=6.5 mm               & 0.9959      &  \ref{eq:pcpf}  \\
\cellcolor[HTML]{EFEFEF}                     & D=7.5 mm               & 0.9988      & \multirow{-2}{*}{} \\
\cellcolor[HTML]{EFEFEF}\textbf{Unsupported} & t=9.5 mm               & 0.9951      &   \ref{eq:pcpf}\\
\cellcolor[HTML]{EFEFEF}                     & t=8.5 mm               & 0.9888      & \multirow{-2}{*}{} \\
\cellcolor[HTML]{EFEFEF}\textbf{Supported}   & t=6.5 mm               & 0.9959      & \ref{eq:pcpf}  \\
\cellcolor[HTML]{EFEFEF}                     & t=7.5 mm               & 0.9988      & \multirow{-2}{*}{} \\
\cellcolor[HTML]{EFEFEF}\textbf{Embossed}    & h=3.5 mm x 3.5 mm       & 1.0000      &  \ref{eq:pcpf}                                             \\
\cellcolor[HTML]{EFEFEF}                     & h=4.5 mm x 4.5 mm       & 1.0000      & \multirow{-2}{*}{} \\
\cellcolor[HTML]{EFEFEF}\textbf{Engraved}    & d=3.5 mm x 3.5 mm       & 1.0000      &  \ref{eq:pcpf}  \\
\cellcolor[HTML]{EFEFEF}                     & d=4.5 mm x 4.5 mm       & 1.0000      & \multirow{-2}{*}{}  \\
\cellcolor[HTML]{EFEFEF}\textbf{Thin parts}  & t=6.5 mm               & 0.9959      & \ref{eq:pcpf}  \\
\cellcolor[HTML]{EFEFEF}                     & t=7.5 mm               & 0.9988      & \multirow{-2}{*}{}\\
\cellcolor[HTML]{EFEFEF}\textbf{Overhangs}   & s=$1.5276E$+$04\,N/m^2$       & 0.7505      &   \ref{eq:pcpf_transf} \\
                                             &                     & \cellcolor[HTML]{E0C2CD}\textbf{$\prod (1-P_{F})$} & \cellcolor[HTML]{E0C2CD}\textbf{0.7255} \\
\rowcolor[HTML]{CBCEFB} 
\textbf{Sensitivity factor}                  & \textbf{1-PG}       & \textbf{P(M,T)}  & \textbf{OPS (\%)}   \\
for k=0.1                                    & 0.99004             & 0.71828          & \textbf{71.82\%}    \\
for k=0.5                                    & 0.95089             & 0.68988          & \textbf{68.98\%}   \\
for k=0.9                                    & 0.91288             & 0.66230          & \textbf{66.23\%}                   
\end{tabular}
}
\caption{Overall printability score calculation for Benchmark 3 (D:diameter, t:thickness, h:height, w:width, d:depth and s:stress) (table by authors).}
\label{tab:benchmark 3-printability}
\end{table}

As shown in Tables \ref{tab:benchmark 1-printability}, \ref{tab:benchmark 2-printability}, and \ref{tab:benchmark 3-printability}, as the value of the part characteristic dimension becomes smaller than the corresponding $w_d(T,i)$ (Table \ref{tab:critical}), the probability of a successful print decreases. In the case of overhang geometries, the dimension values that yield stresses lower than the critical $w_d(T,i)$ lead to a higher $(1-P_F)$ probability.

\begin{figure}[htbp]
 \centering
 \includegraphics[width=0.9\linewidth]{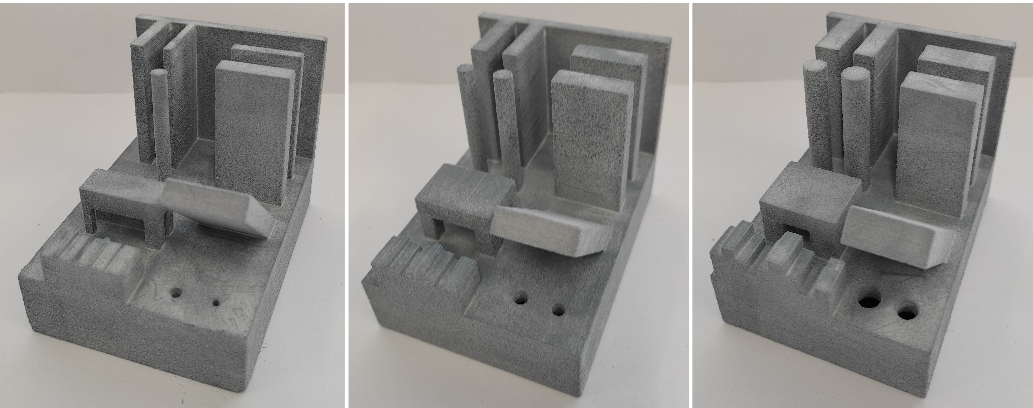}
 \caption{Printed benchmark models: Benchmark 1 (left), Benchmark 2 (center) and Benchmark 3 (right) (figure by authors).}
 \label{fig:bench Print}
\end{figure}

All the benchmark models were printed four times. Benchmark 1, with the lowest overall printability score (mean overall printability score $\approx 3.17\%$, over different sensitivity $k$ values), had the most printing failures because half of the part characteristic dimensions are close to the corresponding $w_d(T,i)$ values. In three out of four manufactured models the pins with $D=2.5\,mm$ broke during depowdering while the thinner unsupported walls ($t=3.5\,mm$) were quite unstable. For Benchmark 2 the same part characteristics for the smaller dimension value have been printed successfully but are quite unstable. Finally, Benchmark 3 does not present any particular problems or failures. These print results are consistent with the overall printability score estimation. The printed models are shown in Figure \ref{fig:bench Print}.

The results of this section have shown that structural robustness for thin parts and overhangs is captured quite accurately by part characteristic probability function of the printability framework. Finally, the global probability function of the printability framework correctly characterizes the quality of the printed part based on the characteristics of the AM technology, the accuracy of the triangulated model and the application domain.

\section{Conclusions}

We have reported on the development of a framework for predicting errors and failures in additive manufacturing. To this end, we  introduced the following:
\begin{itemize}
\item[(i)] A method for predicting dimensional errors that may occur when printing a 3D model. This method was verified using BJ technology. However, it can be extended to other technologies as well, if a similar data acquisition scheme is used to accurately reconstruct the fabricated object. 
\item[(ii)]  An approach to estimate the probability of a CAD model to be fabricated without significant failures or errors that make it unsuitable for a specific application. This approach has been verified for three technologies (FDM, BJ and MJ) but can be easily extended to other technologies as data for the fidelity and robustness of the fabricated objects become available for these technologies as well.
\end{itemize}

A promising research direction is the automated correction of the original CAD model or the corresponding mesh to transform parts with a high probability of failure. In addition, the use of convolutional neural networks may provide even better predictions for dimensional errors when applied to a mesh graph by detecting error-prone topological patterns.

 In the present study, only part characteristics related to structural robustness and detail capture were used to characterize printability. In future work, additional part characteristics such as escape holes and moving/connected parts can be investigated and incorporated into the printability framework.

\section*{Appendix A: Verifying claims $C1$ and $C2$}
To verify $C1$, several models with the same length and width values were created with different thickness values for thin parts and two different length values for several diameter values for pins.
 Corresponding finite element models were created and FE analysis was performed to achieve maximum equivalent stresses (von Mises)  for all cases with thickness and diameter values in the range $1.0 \, mm$ to $5.0 \, mm$. 

For the purposes of the analysis all models were clamped on a base under it's own self weight with acceleration value $g=9.8\,m/s^2$. The 3D printer used was ZPrinter 450 (BJ technology) and the material used was a composition of zp151 powder and z63 binder with measured density equal to $1360\,kg/m^3$ before glue infiltration. All models were meshed with linear tetrahedral elements.

\begin{figure}[htpb]
  \begin{subfigure}[c]{\columnwidth}
   \centering
    \includegraphics[width=0.85\linewidth]{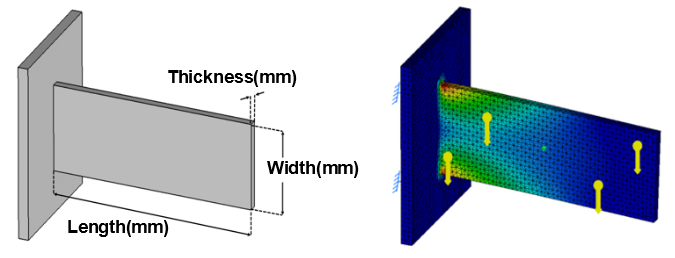}
    \caption{CAD - FE analysis}
    \label{fig:CAD - FE thin part}
  \end{subfigure}
  \begin{subfigure}[c]{\columnwidth}
  \centering
    \includegraphics[width=0.75\linewidth]{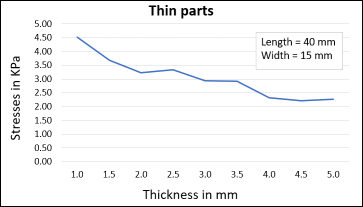}
    \caption{Maximum equivalent stresses}
    \label{fig:stress thin part}
  \end{subfigure}
\caption{Mechanical behaviour representation for thin parts (figure by authors).}
\label{fig:Mechanical behaviour thin}
\end{figure}

In Figure \ref{fig:Mechanical behaviour thin} the CAD model and FE analysis as well as the maximum stresses are presented for the case of thin parts. The length and width values were set to $40\,mm$ and $15\,mm$ correspondingly. Thickness values lie within $1.0$ to $5.0\,mm$ with a step of $0.5\,mm$. The  maximum  stresses  are maximized ($4.52\,kPa$ for $1.0\,mm$ thickness) leading to fracture and printing failure. As the thickness increases, we obtain a more robust construction as maximum stresses decrease ($2.26\,kPa$ for $5.0\,mm$ thickness).

\begin{figure}[htpb]
  \begin{subfigure}[c]{\columnwidth}
   \centering
    \includegraphics[width=0.8\linewidth]{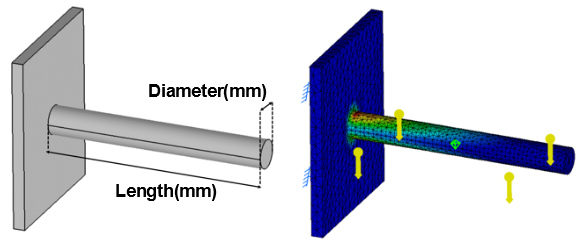}
    \caption{CAD - FE analysis}
    \label{fig:CAD - FE Model pin}
  \end{subfigure}
  \begin{subfigure}[c]{\columnwidth}
  \centering
    \includegraphics[scale=0.7]{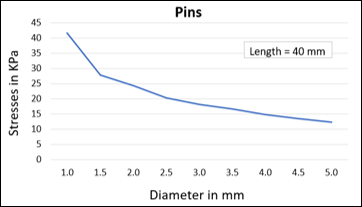}
    \caption{Maximum equivalent stresses for length 40 mm}
    \label{fig:stress pin 40}
  \end{subfigure}
  \begin{subfigure}[c]{\columnwidth}
  \centering
    \includegraphics[scale=0.7]{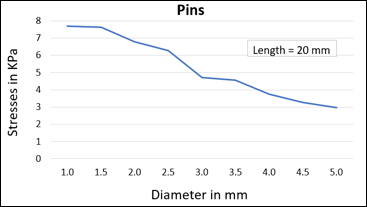}
    \caption{Maximum equivalent stresses for length 20 mm}
    \label{fig:stress pin 20}
  \end{subfigure}
\caption{Mechanical behaviour representation for pins (figure by authors).}
\label{fig:Mechanical behaviour pin}
\end{figure}

Similar results are presented for pins in Figure \ref{fig:Mechanical behaviour pin}. Two models were tested under the same diameter values in $[1.0\,mm, 5.0\,mm]$ with step$=0.5\,mm$ but with different lengths $40\,mm$ and $20\,mm$. The results show small maximum stress values for large diameters and large maximum stress values for small diameters with $length=40\,mm$, maximum stress $41.6kPa$ for $1.0\,mm$ diameter and minimum stress $12.3\,kPA$ for $5.0\,mm$ diameter. For length$=20\,mm$ again an almost linear behaviour is observed with maximum stress $7.69\,kPa$ for the smaller diameter value=$1.0\,mm$  and minimum stress $2.97\,kPa$ for diameter$=5.0\,mm$.

\begin{figure*}
    \begin{subfigure}[b]{1\textwidth}
        \centering
        \includegraphics[scale=0.40]{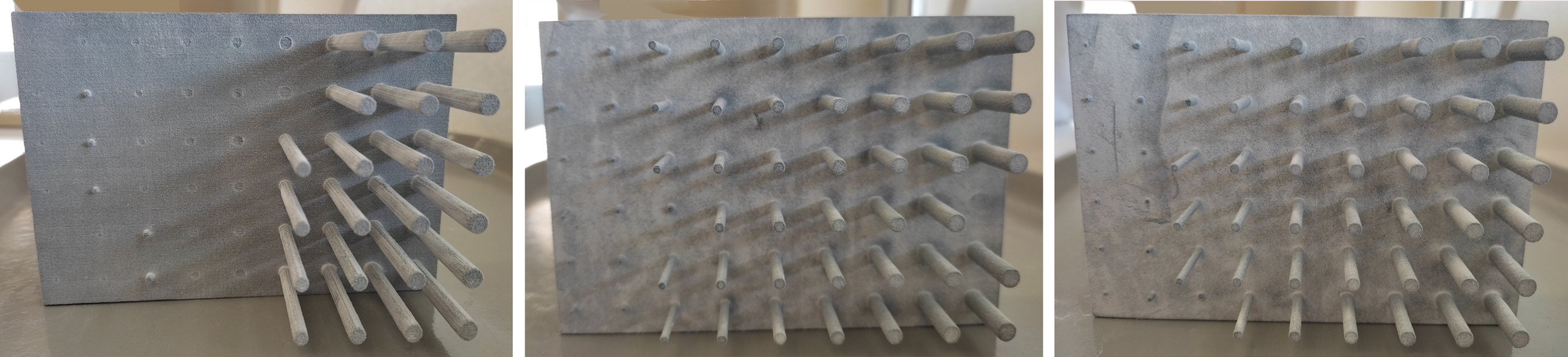}
        \caption{Benchmarks for part characteristic "pins"}
        \label{fig:pin_bench}
    \end{subfigure}
    \begin{subfigure}[b]{1\textwidth}
        \centering
        \includegraphics[scale=0.40]{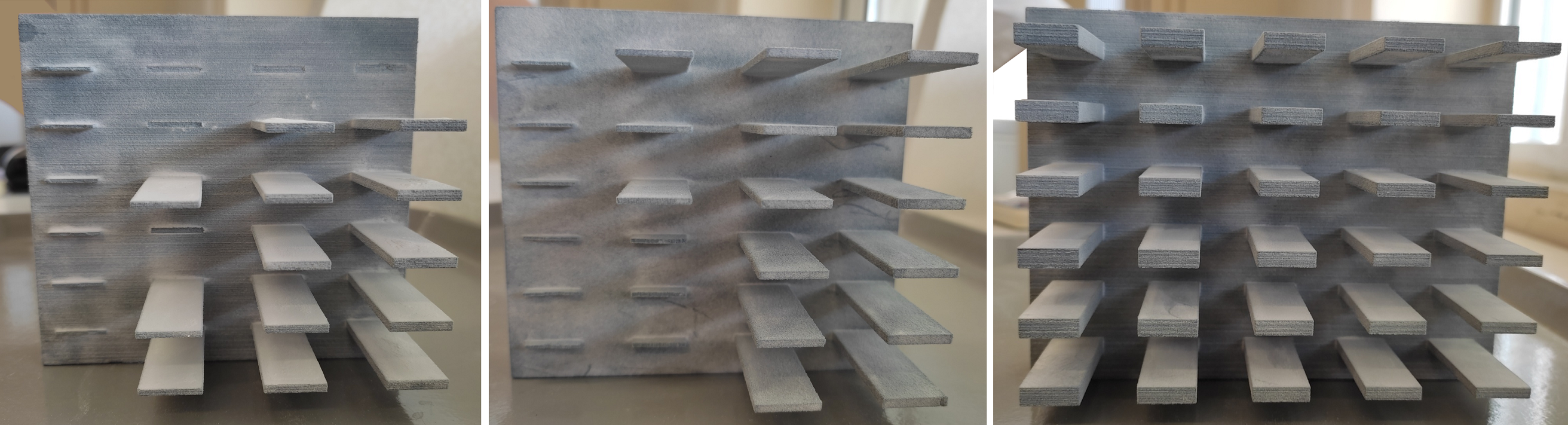}
        \caption{Benchmarks for part characteristic "thin parts"}
        \label{fig:thin bench}
    \end{subfigure}
\caption{Printed benchmarks for printability evaluation (figure by authors).}
\label{fig:printed linearity bench}
\end{figure*}

To verify claim $C2$ all models were printed and the robustness of the fabricated part was characterized. Two different models were printed for the part characteristic ``pins". The first one was printed once for length$=40\,mm$ (Figure \ref{fig:pin_bench} left) and the second four times for length$=20\,mm$ (Figure \ref{fig:pin_bench} middle and right). Six pins were fabricated for each diameter in $[1.00\,mm, 5.00\,mm]$. Based on the manufacturers, the critical value $w_d{(T,i)}$ was set to $2.0\,mm$ with 50\% probability of failure. The length of the pin plays an important role since, for $length=40\,mm$, 6/6 pins with diameter in $[1.0\,mm, 3.0\,mm]$ broke during the depowdering process inside the printer ($100\%$ failure) while, for diameter $3.5\,mm$, 2 out of 6 broke ($33.33\%$ failure) but the remaining are very unstable. For the second model, with a length of $20\,mm$, 24/24 pins with diameters $1.0\,mm$ and $1.5\,mm$ broke ($100\%$ failure) during depowdering. For diameters $2.0\,mm$ and $2.5\,mm$, 13/24 ($54.16\%$ failure) and 4/24 ($16.67\%$ failure) broke during post processing. For diameters $3.0\,mm$ and $3.5\,mm$,  only 1 out of 24 broke ($4.17\%$ failure) during post processing. The remaining pins are stable and robust. Is is observed that the results of pins with length=$20\,mm$ are closer to the constraints provided by the manufacturers for the given critical value $w_d{(T,i)}=2.0\,mm$ for model failure. 

Two more models were printed to evaluate the thickness of the thin parts (Figure \ref{fig:thin bench}). The first benchmark contained the smaller thickness values in $[1.0\,mm, 2.5\,mm]$ and was printed four times (Figure \ref{fig:thin bench} left and middle). For thickness $1.0\,mm$ and $1.5\,mm$, 24/24 pins ($100\%$ failure) and 9/24 pins ($37.5\%$ failure) broke correspondingly inside the printer during depowdering, while two more were very unstable. For the critical value provided by the manufacturers ($w_d{(T,i)}=2.0\,mm$), 3/24 broke ($12.5\%$ failure) but three of the remaining where very unstable. Finally, 2 out of 24 broke ($8.33\%$ failure) for $2.5\,mm$ during post processing. Benchmark 2 (Figure \ref{fig:thin bench} right) was printed only once because it has $0\%$ printing failure for all thickness values in $[3.0\,mm, 5.0\,mm]$ and the complete construction is very robust.

\begin{figure}[htpb]
  \begin{subfigure}[c]{\columnwidth}
   \centering
    \includegraphics[width=0.95\linewidth]{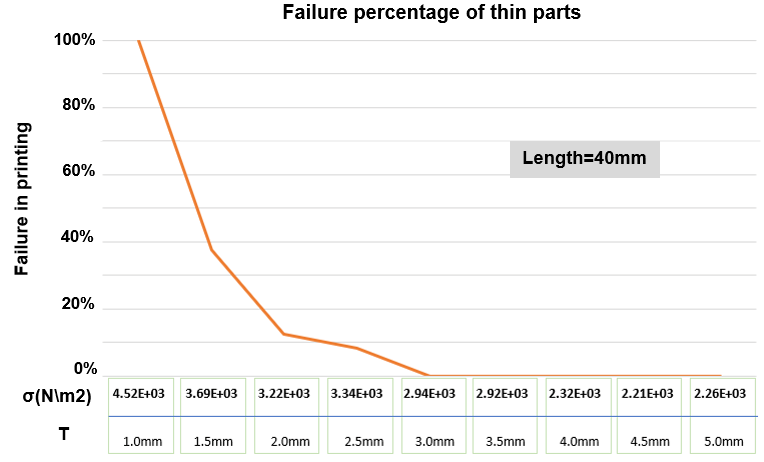}
    \caption{Failure percentage due to maximum equivalent stresses for thin parts}
    \label{fig:failure thin}
  \end{subfigure}
  \begin{subfigure}[c]{\columnwidth}
   \centering
    \includegraphics[width=0.95\linewidth]{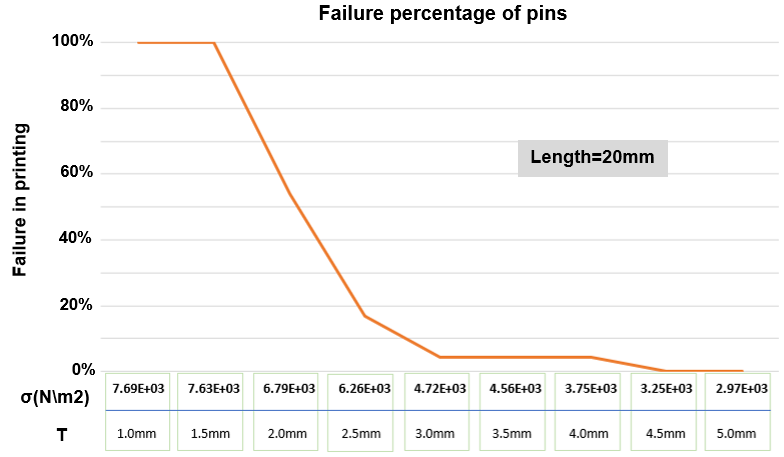}
    \caption{Failure percentage due to maximum equivalent stresses for pins}
    \label{fig:failure pin}
  \end{subfigure}
\caption{Association between failure and maximum equivalent stresses (figure by authors).}
\label{fig:failure}
\end{figure}

Figures \ref{fig:failure thin} and \ref{fig:failure pin} illustrate the relationship between the maximum stresses $\sigma$ (for different thickness values $T$) and failure probability for "thin parts" and "pins" respectively. As shown in both diagrams the maximum percentage of failure was observed for maximum stresses and decreased almost linearly as the maximum stresses decreased.

Since we perform a simple static linear analysis the resulting stresses are affected of the material properties, the applied forces and model geometry. The stress is the ratio of the applied forces over the cross-sectional area of the model. Since the same material is used in all cases and the applied forces are gravitational the area of the model plays the most significant role in reducing or maximizing the defined stresses. A higher value in thickness parameter under the same parameters of length and width will lead to a reduction of this ratio, also by maintaining a consistency in the type and number of the elements, forming the final meshed model, a more uniform distribution of stresses is achieved leading to failure percentage reduction.  

We have shown that the results of the analysis performed for the two part characteristics are consistent with those of the corresponding experiments thus verifying the correctness of the printability framework presented in Section \ref{sec:printability parameters}.
\\
{\bf Acknowledgements:}  This research has been co-financed by the European Union and Greek national funds through the operational Program Competitiveness, Entrepreneurship and Innovation, under the call RESEARCH-CREATE-INNOVATE (project code: T1EDK04928)


\bibliographystyle{elsarticle-harv}
\bibliography{refs}

\end{document}